\pdfoutput=1
\documentclass[aps,prd,amsmath,floats,floatfix, twocolumn,
superscriptaddress,nofootinbib,showpacs,longbibliography]{revtex4-1}
\usepackage[T1]{fontenc}
\usepackage[utf8]{inputenc}
\usepackage{lmodern}
\usepackage{verbatim}
\usepackage[dvipsnames, usenames]{xcolor}
\definecolor{linkcolor}{rgb}{0.0,0.3,0.5}
\usepackage[hypertexnames=false, unicode, colorlinks=true, linkcolor=linkcolor,
citecolor=linkcolor, filecolor=linkcolor,urlcolor=linkcolor,
pdfusetitle]{hyperref}
\usepackage[all]{hypcap}
\usepackage{graphicx}
\usepackage{xspace}
\usepackage{amssymb}
\usepackage{microtype}

\usepackage[english]{babel}
\usepackage{blindtext}

\usepackage[normalem]{ulem} 
\usepackage{bm} 
\usepackage{mathrsfs}

\graphicspath{
	{plots/}
}

\DeclareMathAlphabet{\mathpzc}{OT1}{pzc}{m}{it}

\usepackage[nomessages]{fp}

\begin{document}
\title{Mapping between black-hole perturbation theory and numerical relativity: gravitational-wave energy flux}
\newcommand{\UMassDMath}{\affiliation{Department of Mathematics,
		University of Massachusetts, Dartmouth, MA 02747, USA}}
\newcommand{\UMassDPhy}{\affiliation{Department of Physics,
		University of Massachusetts, Dartmouth, MA 02747, USA}}
\newcommand{\CSCVRUMass}{\affiliation{Center for Scientific Computing and Data Science Research, University of Massachusetts, Dartmouth, MA 02747, USA}}

\author{Tousif Islam}
\email{tislam@umassd.edu}
\UMassDPhy
\UMassDMath
\CSCVRUMass

\hypersetup{pdfauthor={Islam et al.}}

\date{\today}

\begin{abstract}
We investigate the $\alpha$-$\beta$ mapping, as previously introduced by Islam et al.~\cite{Islam:2022laz}, which relates numerical relativity (NR) and adiabatic point-particle black hole perturbation theory (BHPT) waveforms in the comparable mass regime for quasi-circular, non-spinning binary black holes. This mapping involves scaling the amplitude of individual modes with different values of $\alpha$ and the time (and therefore the phase) with a single parameter, $\beta$. In this paper, we demonstrate that this scaling, both in terms of time and orbital frequencies, also extends to the overall gravitational-wave energy flux. This means that we can find a single $\alpha_{\mathcal{F}}$ that scales the BHPT flux and a single $\beta_{\mathcal{F}}$ (which matches the value of $\beta$) that scales the BHPT time such a way that it aligns with NR flux evolution. We then explore the connection between the scaling parameter $\alpha_{\mathcal{F}}$ ($\beta_{\mathcal{F}}$) and the missing finite size correction for the secondary black hole within the BHPT framework.
\end{abstract}

\maketitle

\section{Introduction}
Understanding the interaction between numerical relativity (NR)~\cite{Mroue:2013xna,Boyle:2019kee,Healy:2017psd,Healy:2019jyf,Healy:2020vre,Healy:2022wdn,Jani:2016wkt,Hamilton:2023qkv} and black-hole perturbation theory (BHPT)~\cite{Sundararajan:2007jg,Sundararajan:2008zm,Sundararajan:2010sr,Zenginoglu:2011zz,Fujita:2004rb,Fujita:2005kng,Mano:1996vt,throwe2010high,OSullivan:2014ywd,Drasco:2005kz} is one of the most exciting avenues~\cite{Lousto:2010tb,Lousto:2010qx,Nakano:2011pb,NavarroAlbalat:2022tvh,Albalat:2022lfz,Ramos-Buades:2022lgf,vandeMeent:2020xgc,LeTiec:2014oez,LeTiec:2013uey,LeTiec:2011dp,LeTiec:2011ru} in gravitational wave research. It offers valuable insights into the respective domains of applicability for each framework and contributes to expanding our understanding of binary black hole (BBH) dynamics in the strong-field regime.

NR simulates a BBH merger by numerically solving the Einstein equations without making any approximations. This approach has been refined over the years for modeling BBH mergers with comparable masses, typically within the range $1 \leq q \leq 10$, where $q:=m_1/m_2$ denotes the mass ratio of the binary, with $m_1$ and $m_2$ representing the masses of the larger and smaller black holes, respectively. In contrast, the point particle BHPT framework assumes that the smaller black hole behaves as a point particle and incorporates only adiabatic terms in its calculations~\cite{Sundararajan:2007jg,Sundararajan:2008zm,Sundararajan:2010sr,Zenginoglu:2011zz,Fujita:2004rb,Fujita:2005kng,Mano:1996vt,throwe2010high,OSullivan:2014ywd,Drasco:2005kz}. This means that BHPT waveforms are accurate primarily in the extreme mass ratio limit, where $q$ approaches infinity.
As the binary becomes less asymmetric and enters the regime of comparable masses the assumptions of the BHPT framework start to break down. Consequently, it becomes less reliable in generating accurate gravitational waveforms in this regime. Conversely, as the binary transitions into the intermediate mass ratio regime ($10 \leq q \leq 100$), NR encounters challenges in accurately simulating BBH mergers due to the increasing algorithmic complexity involved. Unsurprisingly, a lot of recent efforts are focused in pushing the regime of validity of each framework. These works involve developing a fully second-order self-force waveforms for non-spinning binary black holes~\cite{Pound:2021qin,Miller:2020bft,Wardell:2021fyy}, discovery of a simple underlying relation named the $\alpha$-$\beta$ mapping between BHPT and NR waveforms in the comparable mass regime~\cite{Islam:2022laz,Rifat:2019ltp}, building NR-tuned BHPT waveform model for comparable to extreme mass ratio binary~\cite{Islam:2022laz,Islam:2023mob,Islam:2023qyt,Islam:2023aec,Islam:2023jak} and performing the first set of high mass ratio NR simulations up to $\sim q=1000$~\cite{Lousto:2020tnb, Lousto:2022hoq}.

The $\alpha$-$\beta$ mapping between the BHPT and NR waveforms reads~\cite{Islam:2022laz}:
\begin{align} \label{eq:EMRI_rescale}
h^{\ell,m}_{\tt NR}(t_{\tt NR} ; q) \sim {\alpha_{\ell}} h^{\ell,m}_{\tt ppBHPT}\left( \beta  t_{\tt ppBHPT};q \right) \,,
\end{align}
where, $h^{\ell,m}_{\tt NR}$ and $h^{\ell,m}_{\tt BHPT}$ represent the NR and BHPT waveforms, respectively, as functions of the NR time $t_{\tt NR}$ and BHPT time $t_{\tt BHPT}$ respectively. Following the $\alpha$-$\beta$ mapping, the scaled BHPT waveform exhibits an excellent agreement with NR in the comparable mass regime, with errors of approximately $10^{-3}$ or less in the quadrupolar mode~\cite{Islam:2022laz}. Additionally, the scaled BHPT waveforms demonstrate a remarkable match to recent high mass ratio ($q=15$ to $q=128$) NR data~\cite{Islam:2023qyt}. Further analysis shows clear evidence that the calibration parameters $\alpha_\ell$ and $\beta$ are related to the absence of finite size of the secondary in the BHPT framework~\cite{Islam:2023aec}.

In this paper, we explore the possibility of such a straightforward mapping, akin to 
the $\alpha$-$\beta$ scaling observed between NR and BHPT waveforms, for the gravitational-wave 
fluxes calculated using NR and BHPT. To gain insights into this, we have examined the relationships between NR and BHPT  waveforms and fluxes for quasi-circular, non-spinning binary systems with mass ratios spanning from 
$q=3$ to $q=10$. We report our findings in Section~\ref{sec:mapping}. In particular, we state the overall flux mapping in Section~\ref{sec:alpha_beta_flux}, demonstrate the effectiveness of the mapping in Section~\ref{sec:q4} and present the functional form of the scaling parameters in Section~\ref{sec:mass_ratio_dependence}. We then discuss the relation between the flux mapping and the missing finite size effect in BHPT in Section~\ref{sec:finite_size}. Finally, we outline the implications and limitations of our results in Section~\ref{sec:understanding_mapping}.

\begin{figure*}
\includegraphics[width=\textwidth]{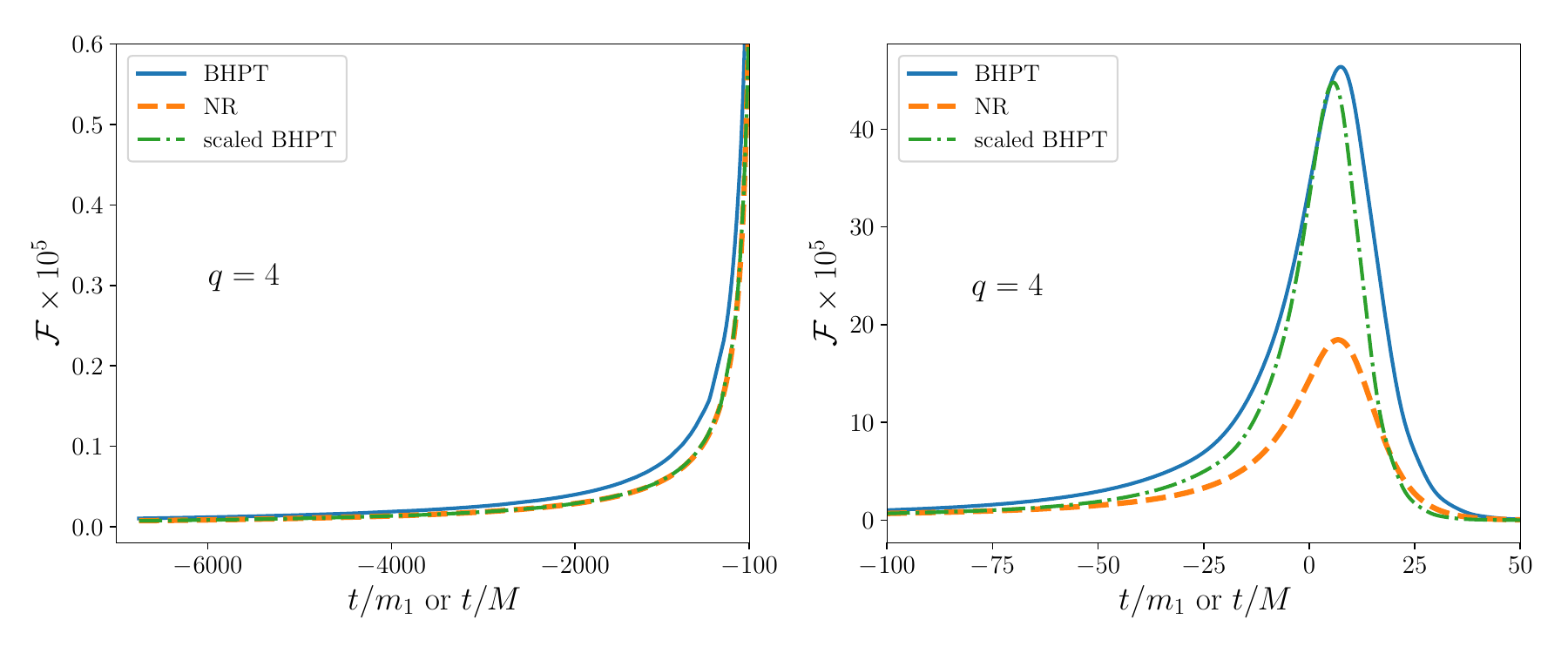}
\caption{We show the fluxes obtained from BHPT (solid blue lines) and NR (dashed orange lines) for a binary with mass ratio $q=4$ along with $\alpha_{\mathcal{F}}$-$\beta_{\mathcal{F}}$ scaled BHPT flux (using Eq.(\ref{eq:flux_scaling})) as green dashed-dotted lines. Left panel shows the inspiral part (where the mapping works extremely well) whereas right panel zooms into the merger-ringdown (where the mapping breaks down due to different final mass/spin scale). More details are in Section \ref{sec:q4}.}
\label{fig:q4_fluxes}
\end{figure*}

\section{Mapping between BHPT and NR fluxes}
\label{sec:mapping}
We utilize the following NR simulations from the SXS collaborations~\cite{Mroue:2013xna,Boyle:2019kee} in our study: $q=3$ (\texttt{SXS:BBH:2265}), $q=3.5$ (\texttt{SXS:BBH:0193}), $q=4$ (\texttt{SXS:BBH:1220}),  $q=4.5$ (\texttt{SXS:BBH:0295}), $q=5.0$ (\texttt{SXS:BBH:0107}), $q=5.5$ (\texttt{SXS:BBH:0296}), $q=6$ (\texttt{SXS:BBH:0181}), $q=6.5$ (\texttt{SXS:BBH:0297}), $q=7$ (\texttt{SXS:BBH:0298}), $q=7.5$ (\texttt{SXS:BBH:0299}), $q=8.0$ (\texttt{SXS:BBH:0063}), $q=8.5$ (\texttt{SXS:BBH:0300}), $q=9.2$ (\texttt{SXS:BBH:1108}), $q=9.5$ (\texttt{SXS:BBH:0302}) and $q=10.0$ (\texttt{SXS:BBH:1107}). The NR data are typically $\sim 4000M$ to $\sim 7000M$ long in duration (where $M:=m_1+m_2$ is the total mass of the binary) with the exception of \texttt{SXS:BBH:0193} which is $\sim 30000M$ long in duration.

We then generate the BHPT waveforms for these mass ratios using the \texttt{BHPTNRSur1dq1e4} model. \texttt{BHPTNRSur1dq1e4}~\cite{Islam:2022laz} is a reduced-order surrogate model trained on waveform data generated with BHPT framework. The full inspiral-merger-ringdown (IMR) BHPT waveform training data is computed using a time-domain Teukolsky equation solver, the details of which have appeared in the literature extensively~\cite{Sundararajan:2007jg,Sundararajan:2008zm,Sundararajan:2010sr,Zenginoglu:2011zz}. This step benefits from tools (such as \texttt{GremlinEq}~\cite{OSullivan:2014ywd,Drasco:2005kz} and \texttt{BHPTNRSurrogate(s)~\cite{BHPTSurrogate}}) available in the \texttt{Black Hole Perturbation Toolkit}~\cite{BHPToolkit}.

\subsection{$\alpha$-$\beta$ scaling of the flux}
\label{sec:alpha_beta_flux}
Our investigation strongly suggests the presence of a straightforward mapping between NR and BHPT 
fluxes, which takes the following form:
\begin{equation}
    \mathcal{F}_{\tt NR} (t_{\tt NR}) =  \alpha_{\mathcal{F}} \times \mathcal{F}_{\tt BHPT} (\beta_{\mathcal{F}} \times t_{\tt BHPT}),
    \label{eq:flux_scaling}
\end{equation}
where $\mathcal{F}_{\tt NR}$ and $\mathcal{F}_{\tt BHPT}$ are the gravitational-wave 
fluxes calculated using NR and BHPT respectively. Furthermore, the mass-scale of the NR fluxes (and waveforms) are the total mass of the binary $M$ whereas the mass-scale of the BHPT fluxes (and waveforms) are $m_1$. The scaling parameters are $\alpha_{\mathcal{F}}$
and $\beta_{\mathcal{F}}$. The energy fluxes due to gravitational radiation are given by
\begin{align}
\mathcal{F}_{\tt NR} (t_{\tt NR}) = \lim_{r \rightarrow \infty} \frac{r^2}{16\,\pi}
\sum_{\ell, m} \,\left| \frac{\partial  h^{\ell m}_{\tt NR}}{\partial t_{\tt NR}} \right|^2 \; ,
\label{eq:nr_flux}
\end{align}
and
\begin{align}
\mathcal{F}_{\tt BHPT} (t_{\tt BHPT})  = \lim_{r \rightarrow \infty} \frac{r^2}{16\,\pi}
\sum_{\ell, m} \,\left| \frac{\partial  h^{\ell m}_{\tt BHPT}}{\partial t_{\tt BHPT}} \right|^2 \;.
\label{eq:bhpt_flux}
\end{align}
Unless otherwise mentioned, to compute the energy flux~\footnote{We use \texttt{gw\_remnant}~\cite{Islam:2023mob,gwremnant} package to compute the fluxes.}, we use all available NR-tuned modes up to $\ell=5$ in the \texttt{BHPTNRSur1dq1e4} model. We also use $G=c=1$. Furthermore, we set the luminosity distance $r=1$. 

We can also re-write the mapping as a function of the respective instantaneous orbital frequencies as:
\begin{equation}
    \mathcal{F}_{\tt NR} (\omega_{\tt NR}) =  \alpha_{\mathcal{F}} \times \mathcal{F}_{\tt BHPT} \left( \frac{\omega_{\tt BHPT}}{\beta_{\mathcal{F}}}\right),
    \label{eq:scaling_fd}
\end{equation}
where
\begin{align}
\omega_{\tt NR}=\frac{d\phi_{\tt orb, NR}}{t_{\tt NR}},\notag \\
\omega_{\tt BHPT}=\frac{d\phi_{\tt orb, BHPT}}{t_{\tt BHPT}},
\end{align}
with $\phi_{\tt NR}$ and $\phi_{\tt BHPT}$ being the orbital phase of the NR and BHPT waveforms respectively. We compute the orbital phases from the $(2,2)$ mode phases as:
\begin{align}
\phi_{\tt orb, NR} = \phi_{\tt NR}^{22} / 2,\notag \\
\phi_{\tt orb, BHPT} = \phi_{\tt ppBHPT}^{22} / 2.
\end{align}
From Eq.(\ref{eq:scaling_fd}), it becomes evident that the flux scaling is actually a frequency-dependent scaling just like the $\alpha$-$\beta$ scaling of the waveforms~\cite{Islam:2023jak}.

In the subsequent subsections, we demonstrate this mapping for binaries with varying mass ratio values and discuss potential interpretations and subtleties.

\subsection{Demonstration at $q=4$}
\label{sec:q4}
We first focus on a binary with a mass ratio of $q=4$. This choice is optimal for 
examining the interaction between NR and BHPT in the comparable mass regime for two 
key reasons: (i) the binary deviates slightly from the equal mass case, ensuring that our 
results are not limited to special scenarios, and (ii) the mass ratio is not high enough 
to render the disparities between NR and BHPT negligible.

\subsubsection{Effectiveness of the scaling}
\label{sec:effectiveness_alpha_beta}
Figure \ref{fig:q4_fluxes} shows the flux obtained from BHPT (solid blue line) and NR (dashed orange line). For comparison, we also show the $\alpha_{\mathcal{F}}$-$\beta_{\mathcal{F}}$ scaled BHPT flux as green dashed-dotted lines. As expected, the BHPT and NR fluxes do not align, given they are scaled with different mass parameters: $m_1$ for BHPT and $M$ for NR. It is important to emphasize that even with fixed mass scaling, substantial differences between NR and BHPT fluxes persist. However, when we appropriately scale the BHPT fluxes using the values of $\alpha_{\mathcal{F}}$ and $\beta_{\mathcal{F}}$, the scaled BHPT fluxes match the NR fluxes very well. 

Notably, the scaling seems to break down (as observed in Figure \ref{fig:q4_fluxes}, right panel) very close to the merger, occurring approximately $75M$ before the merger~\footnote{We denote the time corresponding to the maximum amplitude of the $(2,2)$ mode as the time of merger.}. The breakdown of the $\alpha_{\mathcal{F}}$-$\beta_{\mathcal{F}}$ scaling for the flux is likely related to the changes in mass and spin values that occur after the plunge, akin to the breakdown observed in the $\alpha$-$\beta$ scaling between BHPT and NR waveforms, as shown in Ref.~\cite{Islam:2023aec}.

\begin{figure}
\includegraphics[width=\columnwidth]{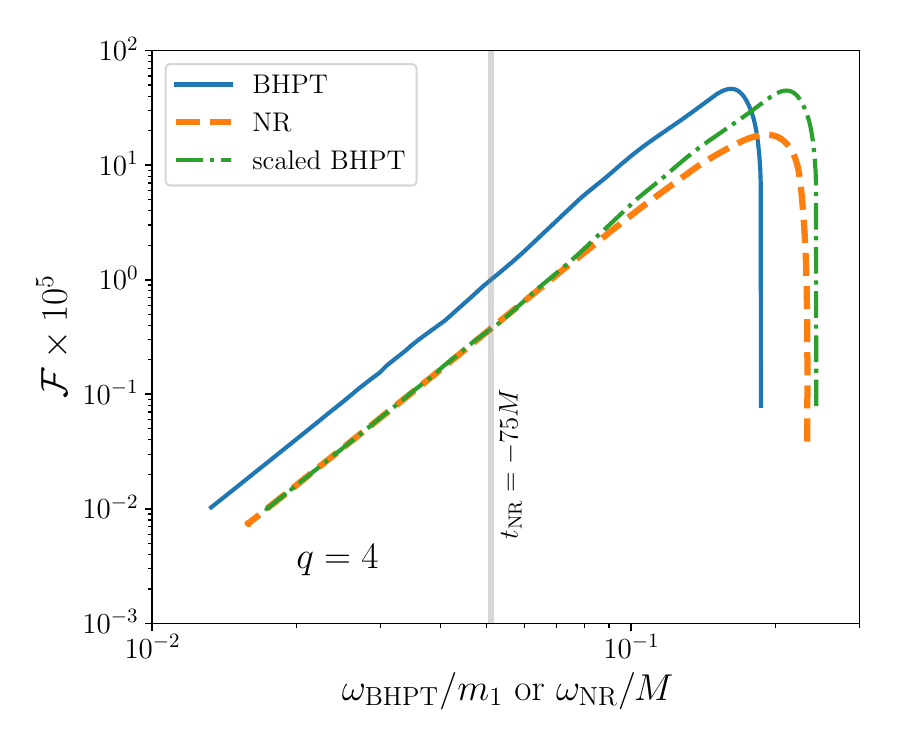}
\caption{We show the fluxes obtained from BHPT (solid blue line) and NR (dashed orange line) for a binary with mass ratio $q=4$ along with the $\alpha_{\mathcal{F}}$-$\beta_{\mathcal{F}}$ scaled BHPT flux (Eq.(\ref{eq:scaling_fd}); as green dashed-dotted line) as a function of their respective orbital frequencies: $\omega_{\rm BHPT}/m_1$ for BHPT and $\omega_{\rm NR}/M$ for NR (and scaled BHPT). More details are in Section \ref{sec:q4}.}
\label{fig:q4_fluxes_omega}
\end{figure}

To illustrate that the $\alpha_{\mathcal{F}}$-$\beta_{\mathcal{F}}$ scaling for the flux represents a frequency-dependent correction, we present the flux obtained from BHPT (solid blue line) and NR (dashed orange line), along with the $\alpha_{\mathcal{F}}$-$\beta_{\mathcal{F}}$ scaled BHPT flux (green dashed-dotted line), plotted against their respective instantaneous orbital frequencies in Figure~\ref{fig:q4_fluxes_omega}. It is evident that the $\alpha_{\mathcal{F}}$-$\beta_{\mathcal{F}}$ scaling simultaneously adjusts the flux and the orbital frequencies to align BHPT fluxes with NR. It further aids in the identification of the breakdown regime more distinctly. Nonetheless, our study demonstrates that the $\alpha_{\mathcal{F}}$-$\beta_{\mathcal{F}}$ scaling performs exceptionally well up to a point very close to the merger. Beyond this point, additional corrections, likely associated with the final mass and spin of the remnant, need to be considered~\cite{Islam:2023mob}. We leave this for future exploration.

\subsubsection{Estimating $\alpha_{\mathcal{F}}$ and $\beta_{\mathcal{F}}$ parameters}
\label{sec:estimate_alpha_beta}
Up to this point, we have primarily showcased the effectiveness of $\alpha_{\mathcal{F}}$-$\beta_{\mathcal{F}}$ scaling throughout the inspiral phase. Now, we provide further insights into how we determine the appropriate values for the $\alpha_{\mathcal{F}}$ and $\beta_{\mathcal{F}}$ parameters. 

First, we make an intelligent guess about the $\beta_{\mathcal{F}}$ parameter. Our earlier analyses~\cite{Islam:2022laz,Islam:2023aec} have unveiled that while scaling BHPT waveforms to match NR, the $\beta$ parameter maintains a consistent value, irrespective of the mode employed for its determination. This consistency is suspected to be related to common post-adiabatic corrections during the binary's evolution. It is therefore more probable that after scaling by the mass-scale transformation factor $\frac{1}{1+1/q}$, $\beta_{\mathcal{F}}$ would also take on the same value as $\beta$. In our notation:
\begin{equation}
    \beta = \frac{1}{1+1/q} \times \beta_{\mathcal{F}}.
\end{equation}
We therefore use the following analytical approximation of $\beta$ (presented in Ref.~\cite{Islam:2022laz}) to obtain its value (and hence the value of $\beta_{\mathcal{F}}$) at $q=4$:
\begin{align}
\begin{split}
\beta(q) =  1 & - \frac{1.238}{q} + \frac{1.596}{q^2} 
- \frac{1.776}{q^3} + \frac{1.0577}{q^4}.\;
\end{split}
\label{beta_fit}
\end{align}

\begin{figure}
\includegraphics[width=\columnwidth]{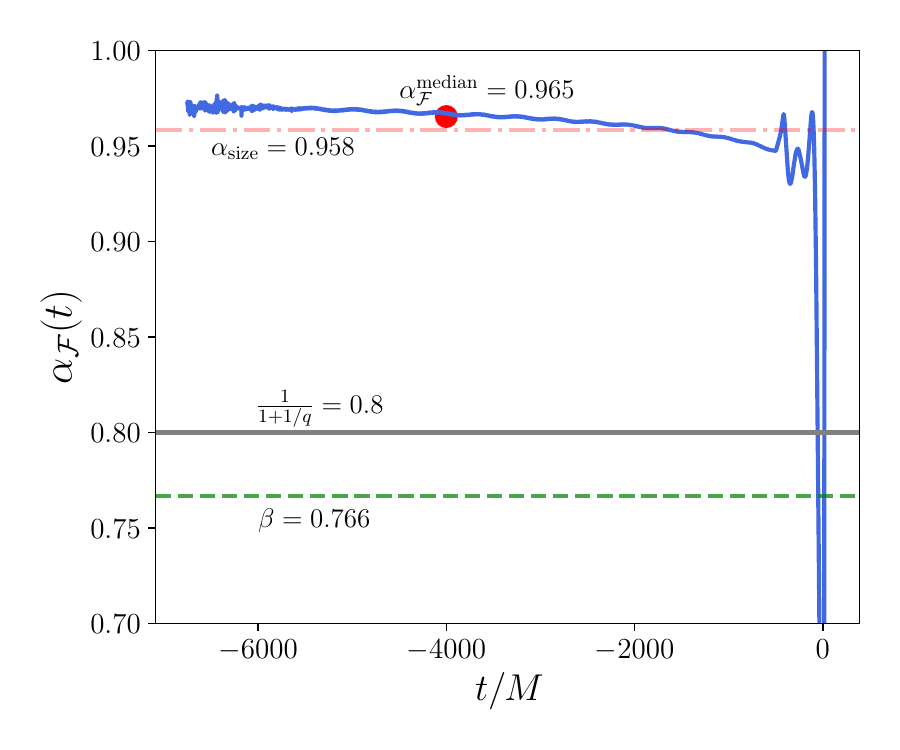}
\caption{We show the temporal variation of the $\alpha_{\mathcal{F}}$ (blue solid line) for a mass ratio of $q=4$. Additionally, we show the naive mass-scale transformation factor $\frac{1}{1+1/q}$ (grey solid line), $\beta$ obtained in Ref.~\cite{Islam:2022laz} (green dashed line) and $\alpha_{\rm size}$ obtained in Ref.~\cite{Islam:2023qyt} (dash-dotted red line). We further show $\alpha_{\mathcal{F}}^{\rm median}$, the median value of $\alpha_{\mathcal{F}}(t)$, as a red circle. More details are in Section \ref{sec:q4}.}
\label{fig:q4_alpha_timeseries}
\end{figure}

Once we scale the BHPT time with $\beta_{\mathcal{F}}$, it effectively compensates for the mass-scale difference. With both the time-rescaled BHPT flux and NR flux now synchronized in time, we calculate the ratio of the time-rescaled BHPT flux to the NR flux to understand the temporal variation of the required $\alpha_{\mathcal{F}}$ parameter (Figure~\ref{fig:q4_alpha_timeseries}): 
\begin{equation}
    \alpha_{\mathcal{F}}(t_{\tt NR}) = \frac{\mathcal{F}_{\tt BHPT} (\beta_{\mathcal{F}} \times t_{\tt BHPT})}{\mathcal{F}_{\tt NR} (t_{\tt NR})}.
    \label{eq:alpha_f_t}
\end{equation}
We observe that the required $\alpha_{\mathcal{F}}$ remains nearly constant up to a point very close to the merger, after which it undergoes rapid changes. We estimate the median value of the required $\alpha_{\mathcal{F}}(t_{\tt NR})$ to be 0.965 (and denote it as $\alpha_{\mathcal{F}}^{\rm median}$) which we use in this section. We also compute an optimized value of $\alpha_{\mathcal{F}}$ for which the difference between scaled BHPT and NR fluxes is the minimum. We denote this to be $\alpha_{\mathcal{F}}^{\rm optimized}$ and find its value to be 0.962 - very close to the value of $\alpha_{\mathcal{F}}^{\rm median}(=0.965)$.

It is important to highlight that the estimated $\alpha_{\mathcal{F}}$ significantly differs from the naive mass-scaling factor (between $m_1$ and $M$) of $\frac{1}{1+1/q}=0.8$, as well as from the value of $\beta(=0.706)$ obtained from the analytical expression mentioned earlier. Furthermore, it diverges from the estimated value of $\alpha_{\ell=2}(=0.766)$. 
Interestingly, the determined value of $\alpha_{\mathcal{F}}^{\rm median}$ closely aligns with the estimated correction parameter $\alpha_{\rm size,\ell=2}(=0.958)$ that accounts for the missing finite size effect of the secondary black hole in BHPT framework~\cite{Islam:2023aec}. The parameter $\alpha_{\rm size,\ell=2}$ is related to the parameter $\alpha_\ell$ as~\cite{Islam:2023aec}:
\begin{equation}
    \alpha_\ell = \frac{1}{1+1/q} \times \alpha_{\rm size,\ell}.
\end{equation}

\subsection{Mass ratio dependence}
\label{sec:mass_ratio_dependence}

We now repeat this analysis for an additional 14 mass ratio values spanning from $q=3$ to $q=10$. Our findings reveal that the $\alpha_{\mathcal{F}}$-$\beta_{\mathcal{F}}$ scaling, as outlined in Section~\ref{sec:alpha_beta_flux}, functions exceptionally well for all these mass ratios. 

\begin{figure*}
\includegraphics[width=\textwidth]{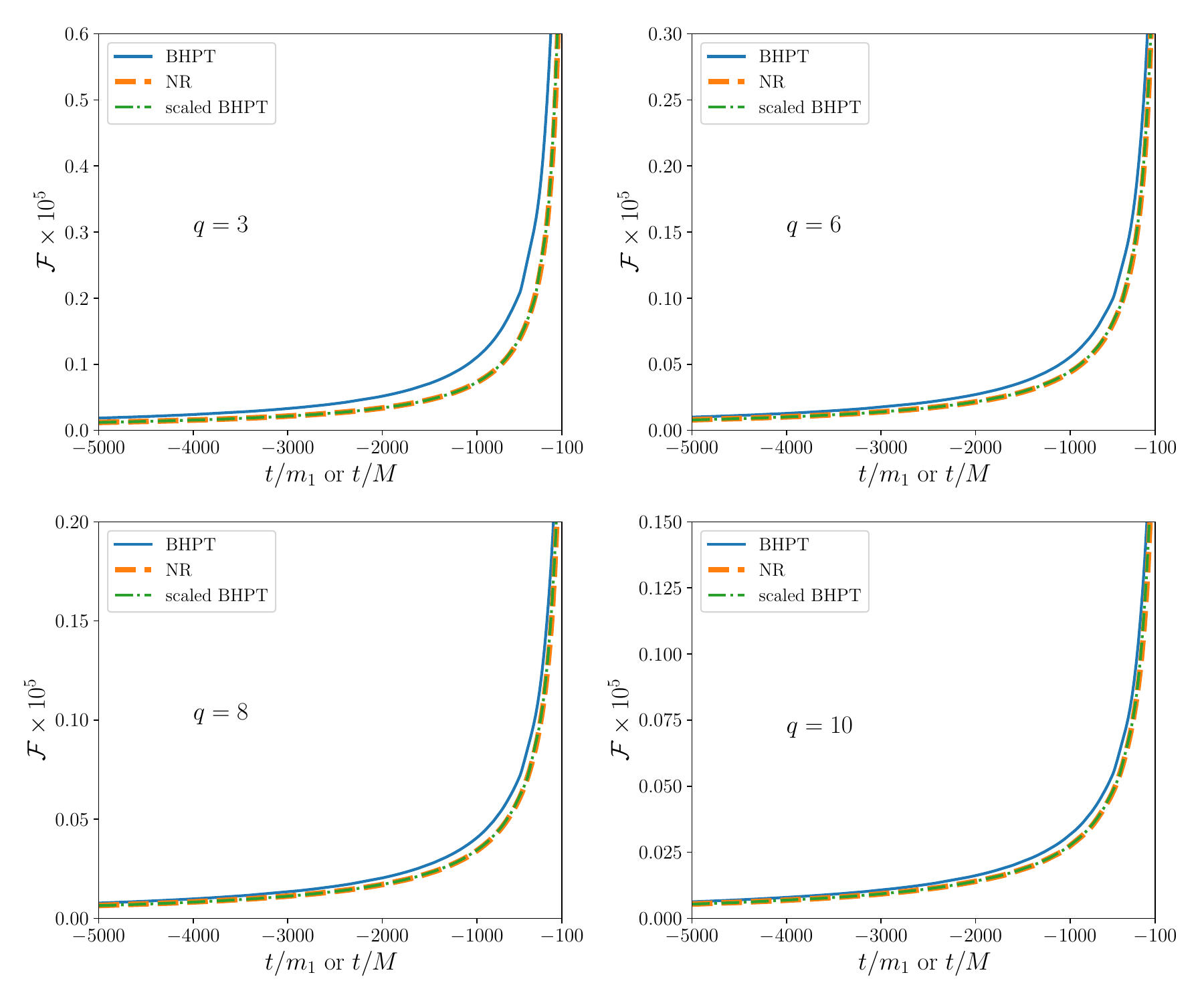}
\caption{We show the fluxes obtained from BHPT (solid blue lines) and NR (dashed orange lines) for a binary with mass ratios ranging from $q=3$ to $q=10$ along with $\alpha_{\mathcal{F}}$-$\beta_{\mathcal{F}}$ scaled BHPT flux (green dashed-dotted lines; Eq.(\ref{eq:flux_scaling})) as a function of their respective times: $t/m_1$ for BHPT and $t/M$ for NR (and scaled BHPT). We only show the inspiral part, up to $t=-100M$, where the mapping works really well. More details are in Section \ref{sec:mass_ratio_dependence}.}
\label{fig:q3_toq10_fluxes}
\end{figure*}

As evidence, Figure ~\ref{fig:q3_toq10_fluxes} presents the flux obtained from BHPT (solid blue lines) and NR (dashed orange lines), alongside the $\alpha_{\mathcal{F}}$-$\beta_{\mathcal{F}}$ scaled BHPT flux (green dashed-dotted lines) for mass ratios $q=[3,6,8,10]$. In each case, while BHPT and NR fluxes do not align, the scaled BHPT flux demonstrates excellent agreement with the NR data. Additionally, it is worth noting that the difference between BHPT and NR flux diminishes as the mass ratio increases. Here, we restrict our focus to the binary's evolution until very close to the merger, as we have previously established (Section~\ref{sec:effectiveness_alpha_beta}) that the scaling breaks down around the time of merger.

\subsubsection{Functional form of $\alpha_{\mathcal{F}}$ and $\beta_{\mathcal{F}}$}
\label{sec:alpha_beta_fit}
Next, we investigate how the value of $\alpha_{\mathcal{F}}$ changes with the mass ratio. 

To ensure meaningful comparisons, we standardize the length of the NR data for all mass ratio values. Specifically, the shortest NR data determines the common length of all NR waveforms used in the analysis. We perform this analysis twice:
\begin{itemize}
    \item First, we consider all 15 NR datasets, ranging from $q=3$ to $q=10$. This constrains the common length of the NR data to approximately $\sim 3300M$. We denote the value of $\alpha_{\mathcal{F}}$ as $\alpha^{\rm median}_{\mathcal{F},3300M}$ to indicate the length of NR data used. 
    \item Subsequently, we repeat the same analysis using only 8 NR datasets with relatively longer duration, for which we use approximately $\sim 5500M$ of data. In this case, we denote the value of $\alpha_{\mathcal{F}}$ as $\alpha^{\rm median}_{\mathcal{F},5500M}$.
\end{itemize}
For both cases, we also compute the value of $\alpha_{\mathcal{F}}^{\rm optimized}$ and denote them as $\alpha_{\mathcal{F},3300M}^{\rm optimized}$ and $\alpha_{\mathcal{F},5500M}^{\rm optimized}$ respectively.

\begin{figure}
\includegraphics[width=\columnwidth]{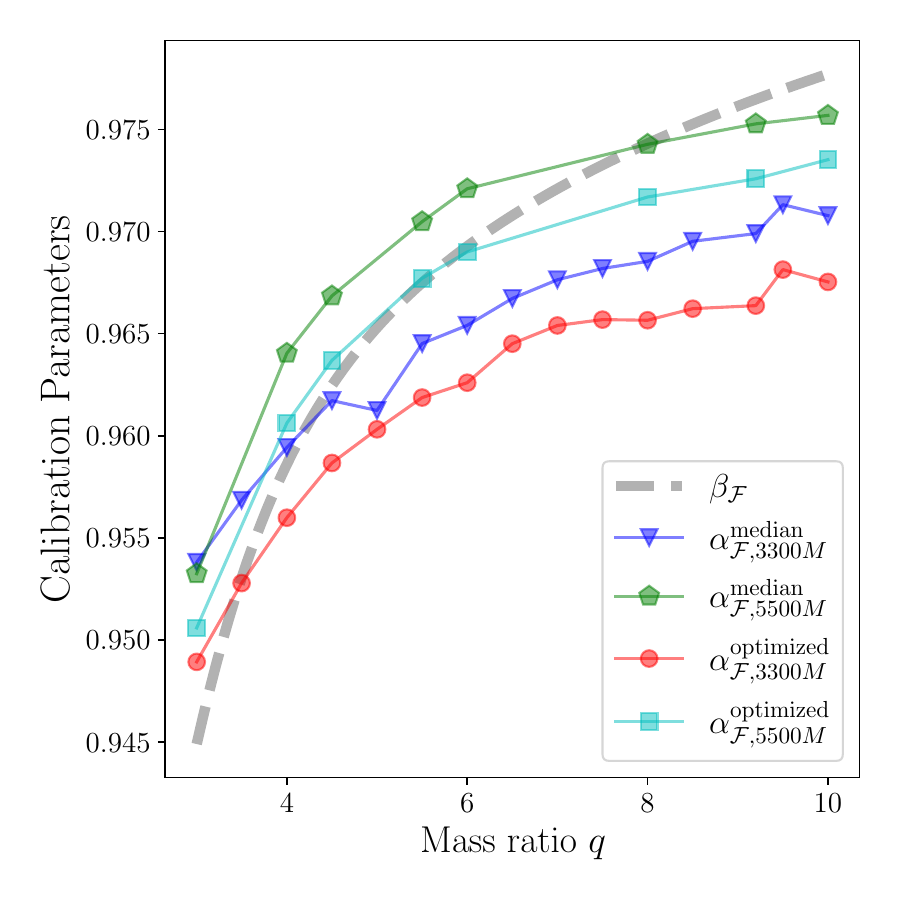}
\caption{We show the calibration parameters $\alpha_{\mathcal{F}}$, obtained using different length of the NR data and methods, and $\beta_{\mathcal{F}}$ as a function of the mass ratio ranging from $q=3$ to $q=10$. More details are in Section \ref{sec:alpha_beta_fit}.}
\label{fig:alpha_flux_vs_q}
\end{figure}

Figure~\ref{fig:alpha_flux_vs_q} illustrates how $\alpha^{\rm median}_{\mathcal{F},3300M}$ and $\alpha^{\rm median}_{\mathcal{F},5500M}$ change with the mass ratio. For reference, we also show the mass ratio dependence of $\beta_{\mathcal{F}}$. It is noteworthy that these values are quite close to each other implying possibly similar origin. Moreover, $\alpha^{\rm median}_{\mathcal{F},5500M}$ tends to take slightly larger values compared to $\alpha^{\rm median}_{\mathcal{F},3300M}$, indicating a subtle distinction between the two based on data length.

Now, we fit $\alpha^{\rm median}_{\mathcal{F},3300M}$ and $\alpha^{\rm median}_{\mathcal{F},5500M}$ in terms of $q$ (using the \texttt{scipy.optimize.curve\_fit}~\cite{scipyfit} module) and obtain:
\begin{align}
\alpha^{\rm median}_{\mathcal{F},3300M} \approx & 0.92966 +  0.011009 \times (\frac{1}{q})\notag\\
&- 0.0010722 \times (\frac{1}{q})^2 + 0.0000385 \times (\frac{1}{q})^3,
\end{align}
\begin{align}
\alpha^{\rm median}_{\mathcal{F},5500M} \approx & 0.893766238 + 0.029565927 \times (\frac{1}{q})\notag\\
&- 0.00365886 \times (\frac{1}{q})^2 + 0.000152464 \times (\frac{1}{q})^3.
\end{align}
These relations provide a crude insight about how the flux correction (through $\alpha_{\mathcal{F}}$ parameters) changes as a function of the mass ratio. Finally, we identify that $\beta_{\mathcal{F}}$ is equivalent to $\beta_{\rm size}$ introduced in Ref.~\cite{Islam:2023aec} which also reports its mass ratio dependence. For the sake of completeness, we provide the functional form below:
\begin{align}
\beta_{\mathcal{F}} \approx & 0.9994668 -  0.2303172 \times (\frac{1}{q})\notag\\
&+0.3294111 \times (\frac{1}{q})^2 - 0.2743849 \times (\frac{1}{q})^3.
\end{align}

\section{Missing finite size effect in BHPT and $\alpha_{\mathcal{F}}$-$\beta_{\mathcal{F}}$ scaling}
\label{sec:finite_size}
It is essential to emphasize that in the BHPT framework, the secondary black hole accurately accounts for its mass but is treated as a point particle, lacking a concept of radius. Conversely, NR simulations incorporate the size of the secondary black hole. This represents one of the most critical distinctions in the treatment between NR and BHPT frameworks, and it consequently results in differences at the waveform and flux levels, particularly in the comparable mass regime where the point particle assumption in BHPT breaks down.

Islam and Khanna~\cite{Islam:2023aec} have recently shown that by treating the secondary black hole as a finite-size object, BHPT waveforms closely align with NR waveforms. They employed a straightforward framework~\cite{Barausse:2021} that models the extended secondary as point particles arranged in a manner consistent with the anticipated shape and size of the secondary black hole. This results in a flux regularization term $f(m)$ such that
\begin{equation}
    \mathcal{F_{\tt tot}} = \sum_{\ell,m} |f(m)|^2 \mathcal{F}^{\ell,m}_{\tt BHPT}
\end{equation}
where $\mathcal{F}^{\ell,m}_{\tt BHPT}$ is the $(\ell, m)$ component of the energy flux from
a single point-like particle of the same total mass-energy in the BHPT framework and $F_{\tt tot}$ is the total flux after the correction. The factor $f(m)$ is given by 
\begin{equation}
  |f(m)|^2=2(1-\cos{\eta})/\eta^2\,,\quad \eta\equiv 2\pi m L/r_0\,.  
\end{equation}
Here $L$ represents the size of the extended object in the azimuthal direction, $m$ is the multipole mode and
$r_0$ is the radius of the circular orbit. This correction is negligible during most of the slow inspiral 
phase of the binary evolution and is stronger around the merger~\cite{Barausse:2021}. Details of the framework is given in Sec. IV of Ref.~\cite{Barausse:2021}. Further discussions on the framework can be found in Ref.~\cite{Islam:2023aec}.

\subsection{Connection between $\alpha_\ell$ and $\alpha_{\mathcal{F}}$ and $f(m)$}
A detailed inspection of Eq.(\ref{eq:EMRI_rescale}), Eq.(\ref{eq:flux_scaling}) and Eq.(\ref{eq:nr_flux}) gives:
\begin{align}\label{eq:flux_to_wf_alpha}
\mathcal{F}_{\tt NR} (t_{\tt NR}) &= \lim_{r \rightarrow \infty} \frac{r^2}{16\,\pi}
\sum_{\ell, m} \,\left| \frac{\partial  (\alpha_\ell h^{\ell m}_{\tt BHPT})}{\partial t_{\tt BHPT}} \frac{\partial t_{\tt NR}}{\partial t_{\tt BHPT}} \right|^2 \notag \\
&= \lim_{r \rightarrow \infty} \frac{r^2}{16\,\pi}
\sum_{\ell, m} \,\left| \alpha_\ell \frac{\partial  h^{\ell m}_{\tt BHPT}}{\partial t_{\tt BHPT}} \frac{1}{\beta}\right|^2 \notag \\
& = \sum_{\ell, m} \,\left| \frac{\alpha_\ell}{\beta}\right|^2 \mathcal{F}_{\tt BHPT}^{\ell,m} .\;
\end{align}
This provides us with an opportunity to establish a connection between $\alpha_\ell$ (and $\beta$) for individual spherical harmonic modes and the approximated flux regularization terms introduced in Ref.~\cite{Barausse:2021}. Within the assumptions of each framework, we can express this as:
\begin{equation}
\frac{\alpha_\ell}{\beta} \approx f(m) = \sqrt{2(1-\cos{\eta})/\eta^2},,\quad
\end{equation}
It is important to note that in the $\alpha$-$\beta$ scaling introduced in Ref.\cite{Islam:2022laz}, for each $(\ell,m)$ mode, $\alpha_\ell$ depends solely on $\ell$ and $\beta$ is mode independent. In contrast, in the flux regularization method introduced in Ref.\cite{Barausse:2021}, $f(m)$ is solely a function of $m$. 
This is, however, unsurprising given that the construction of the scaling factors $f(m)$ and $\alpha_\ell$ has been carried out using $\ell=m$ modes. Both Refs.\cite{Barausse:2021, Islam:2022laz} also noted that the differences in the values of the scaling factors for modes with $\ell \neq m$ compared to $\ell=m$ cases are minimal. Consequently, we can reasonably express $\frac{\alpha(\ell,m)}{\beta}=f(\ell,m)$.

We can go further and can identify that Eq.(\ref{eq:flux_scaling}) implies that it is possible to theoretically compute an overall approximate flux correction factor $f_{\tt tot}$ (in a way similar to Ref.~\cite{Barausse:2021}) such that:
\begin{equation}
    \mathcal{F}_{\tt tot} = f_{\tt tot} \sum_{\ell,m} \mathcal{F}^{\ell,m}_{\tt BHPT}.
\end{equation}
In other words, we can find an overall flux correction factor $f_{\tt tot}$ theoretically such that $\alpha_{\mathcal{F}} = f_{\tt tot}$. This warrants further exploration and is beyond the scope of the current paper.

Finally, using Eq.(\ref{eq:flux_scaling}) and Eq.(\ref{eq:flux_to_wf_alpha}), we can write the following relation between $\alpha_{\mathcal{F}}$ (flux mapping parameter), $\alpha_\ell$ and $\beta$ (waveform mapping parameters) as:
\begin{equation}
\alpha_{\mathcal{F}} \times \mathcal{F}_{\tt BHPT} = \sum_{\ell, m} \,\left| \frac{\alpha_\ell}{\beta}\right|^2 \mathcal{F}_{\tt BHPT}^{\ell,m} .\;
\end{equation}
Simplifying it further, we get
\begin{equation}
\alpha_{\mathcal{F}}  = \frac{\sum_{\ell, m} \,\left| \frac{\alpha_\ell}{\beta}\right|^2 \mathcal{F}_{\tt BHPT}^{\ell,m}}{\mathcal{F}_{\tt BHPT}} .\;
\end{equation}

\begin{figure}
\includegraphics[width=\columnwidth]{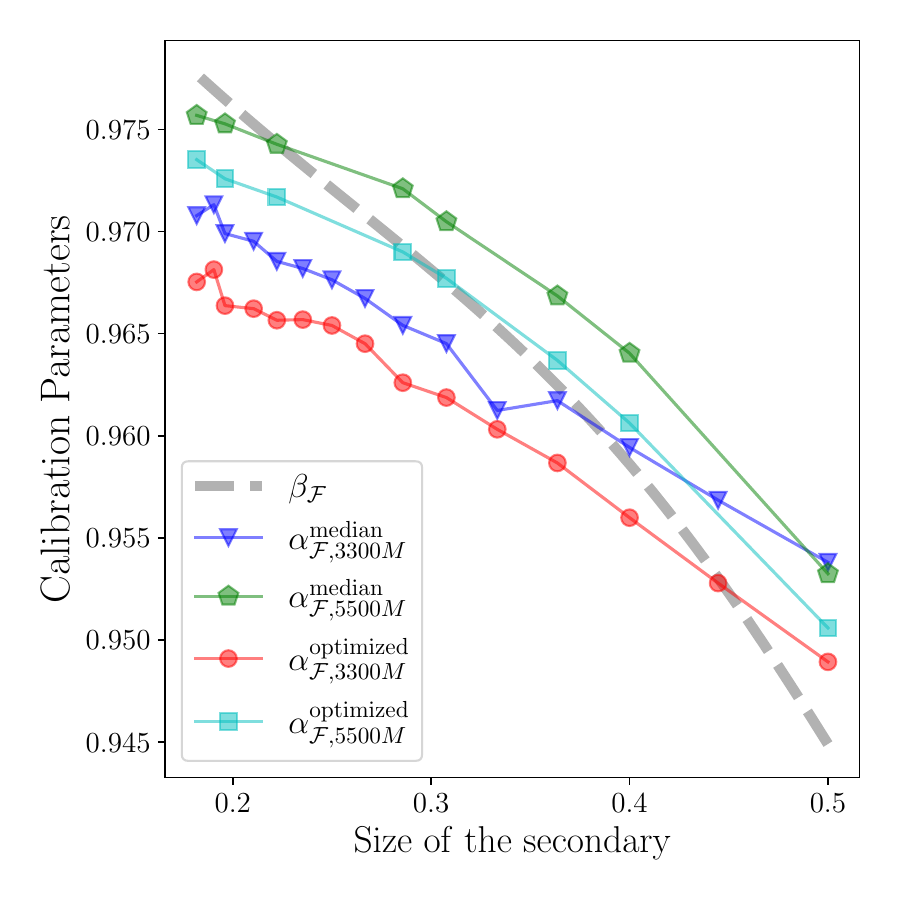}
\caption{We show the calibration parameters $\alpha_{\mathcal{F}}$, obtained using different length of the NR data and methods, and $\beta_{\mathcal{F}}$ as a function of the expected size of the secondary for binaries with mass ratio ranging from $q=3$ to $q=10$. More details are in Section \ref{sec:finite_size}.}
\label{fig:alpha_flux_vs_size}
\end{figure}

\subsection{Connecting $\alpha_{\mathcal{F}}$ to the size of the secondary}
To understand the relation between the missing finite size effect in BHPT and $\alpha_{\mathcal{F}}$-$\beta_{\mathcal{F}}$ scaling, we first compute the horizon area $\mathcal{A}$ of the secondary black hole as
\begin{equation}
\mathcal{A} = 8\pi \left( \frac{Gm_2}{c^2}\right)^2 (1 + \sqrt{1-\chi_2^2}),
\end{equation}
where $\chi_2$ is the spin of the secondary. This enables us to associate a length scale with the black hole through its radius (with $\chi_2=0$ for non-spinning case),
\begin{equation}
r_{\rm S} = \sqrt{\frac{\mathcal{A}}{4\pi}} = \frac{2}{1+q}.
\label{eq:radius}
\end{equation}
In Figure~\ref{fig:alpha_flux_vs_size}, we show how $\alpha^{\rm median}_{\mathcal{F},3300M}$ and $\alpha^{\rm median}_{\mathcal{F},5500M}$ along with $\beta_{\rm \mathcal{F}}$ change with the expected radius of the secondary black hole $r_S$. We then fit $\alpha^{\rm median}_{\mathcal{F},3300M}$ and $\alpha^{\rm median}_{\mathcal{F},5500M}$ in terms of $r_S$ (using the \texttt{scipy.optimize.curve\_fit}~\cite{scipyfit} module) and obtain:
\begin{align}
\alpha^{\rm median}_{\mathcal{F},3300M} \approx & 0.96265727 + 0.19400182 \times r_{\rm S}\notag\\
&-1.20876068 \times r_{\rm S}^2 + 2.49832332 \times r_{\rm S}^3\notag\\
&-1.85580128 \times r_{\rm S}^4
\end{align}
and
\begin{align}
\alpha^{\rm median}_{\mathcal{F},5500M} \approx & 0.98301399 -0.0747335 \times r_{\rm S}\notag\\
&+0.33108185 \times r_{\rm S}^2 - 0.88673433 \times r_{\rm S}^3\notag\\
&+0.57074531 \times r_{\rm S}^4.
\end{align}
From Ref.~\cite{Islam:2023mob}, we already know the functional form of $\beta_{\mathcal{F}} := \beta_{\tt size}$ as: 
\begin{align}
\beta_{\mathcal{F}} \approx & 1.00082016 -  0.1298413 \times r_{\rm S}\notag\\
&+0.07899518 \times r_{\rm S}^2 - 0.05206764 \times r_{\rm S}^3.
\end{align}

\section{Understanding the mapping and future directions}
\label{sec:understanding_mapping}
Now that we have introduced the $\alpha$-$\beta$ scaling between BHPT and NR fluxes and have highlighted its efficacy and limitations across a range of mass ratios within the comparable mass regime, it is crucial to understand the potential implications of our discoveries.

1. First and foremost, it is indeed intriguing that such a straightforward scaling between BHPT and NR fluxes exists. This observation underscores the notion that the linear BHPT framework requires only a minor correction to achieve alignment with fully non-linear NR outcomes.
Our work therefore extends the existence of $\alpha$-$\beta$ scalings from the waveform level~\cite{Islam:2022laz} to the flux level.
    
2. Even more intriguing is the fact that the same value of $\beta$ is effective for both waveform scaling and flux scaling. This provides additional evidence that $\beta$ is associated with overall post-adiabatic corrections that remain consistent across all modes.

3. Another noteworthy observation is that flux provides a more effective means to investigate the interaction between BHPT and NR and to examine the (post-adiabatic) adjustments necessary in the BHPT framework to align with NR. This is because, while time and strain are scaled by the chosen mass-scale (either $m_1$ for BHPT or $M$ for NR), fluxes are dimensionless in geometric units. Consequently, while waveform scale parameter $\alpha$ incorporates mass-scale transformations, flux scale parameter $\alpha_{\mathcal{F}}$ is devoid of mass-scale dependencies.

4. It is also worth noting that the $\alpha_{\mathcal{F}}$ values calculated using different lengths of NR data and various methods exhibit a relatively consistent similarity. Additionally, the value of $\beta_{\mathcal{F}}$ falls within the range of variations in the estimated $\alpha_{\mathcal{F}}$ values. This suggests the possibility that, at the flux level and within the margins of estimation uncertainties, $\alpha_{\mathcal{F}}$ and $\beta_{\mathcal{F}}$ values are identical. Furthermore, this implies that a straightforward flux rescaling with the same values of $\alpha_{\mathcal{F}}$ and $\beta_{\mathcal{F}}$ might enable BHPT to align with NR without requiring any additional calibration, at least until very close to the merger. Such a possibility needs to be investigated further.

This corroborates our earlier work in which we modified the BHPT flux to account for the missing finite size effect~\cite{Islam:2023aec}. We demonstrated that once this modification is implemented, the resulting BHPT waveform aligns remarkably well with NR data. This paper serves as a complementary piece to the findings in the referenced work and together suggest that a straightforward flux correction can effectively address the disparities between BHPT and NR within the comparable mass regime.

5. Lastly, it is crucial to acknowledge the limitations of the scaling we have uncovered. Similar to the $\alpha$-$\beta$ scaling for waveforms, the scaling for flux also exhibits breakdowns very close to the merger. This is likely because the plunge and ringdown phases are more accurately described by the final mass and final spin of the remnant black hole~\cite{Islam:2023mob}. In contrast, BHPT assumes negligible mass and spin changes as the binary evolves, leading to incorrect mass and spin values during the merger and ringdown stages. 

This work serves as a complementary addition to our previous series of studies aimed at unraveling the interplay between NR and the BHPT framework, especially as detailed in Refs.~\cite{Islam:2022laz, Islam:2023aec, Islam:2023mob,Islam:2023jak, Islam:2023qyt}. Collectively, these works offer a comprehensive understanding of the BHPT/NR interplay in case of quasi-circular non-spinning binary black hole mergers. Our future endeavors will include expanding these analyses to encompass precessing and eccentric binary systems.
\begin{acknowledgments}
We thank Gaurav Khanna and Scott Field for helpful discussions and thoughtful comments on the manuscript. We also thank the SXS collaboration for maintaining publicly available catalog of NR simulations which has been used in this study. The author acknowledge support of NSF Grants DMS-1912716.  Simulations were performed on CARNiE at the Center for Scientific Computing and Visualization Research (CSCVR) of UMassD, which is supported by the Naval Research (ONR)/Defense University Research Instrumentation Program (DURIP) Grant No.\ N00014181255 and the UMass-URI UNITY supercomputer supported by the Massachusetts Green High Performance Computing Center (MGHPCC). 
\end{acknowledgments}

\bibliography{References}

\begin{thebibliography}{45}%
\makeatletter
\providecommand \@ifxundefined [1]{%
 \@ifx{#1\undefined}
}%
\providecommand \@ifnum [1]{%
 \ifnum #1\expandafter \@firstoftwo
 \else \expandafter \@secondoftwo
 \fi
}%
\providecommand \@ifx [1]{%
 \ifx #1\expandafter \@firstoftwo
 \else \expandafter \@secondoftwo
 \fi
}%
\providecommand \natexlab [1]{#1}%
\providecommand \enquote  [1]{``#1''}%
\providecommand \bibnamefont  [1]{#1}%
\providecommand \bibfnamefont [1]{#1}%
\providecommand \citenamefont [1]{#1}%
\providecommand \href@noop [0]{\@secondoftwo}%
\providecommand \href [0]{\begingroup \@sanitize@url \@href}%
\providecommand \@href[1]{\@@startlink{#1}\@@href}%
\providecommand \@@href[1]{\endgroup#1\@@endlink}%
\providecommand \@sanitize@url [0]{\catcode `\\12\catcode `\$12\catcode
  `\&12\catcode `\#12\catcode `\^12\catcode `\_12\catcode `\%12\relax}%
\providecommand \@@startlink[1]{}%
\providecommand \@@endlink[0]{}%
\providecommand \url  [0]{\begingroup\@sanitize@url \@url }%
\providecommand \@url [1]{\endgroup\@href {#1}{\urlprefix }}%
\providecommand \urlprefix  [0]{URL }%
\providecommand \Eprint [0]{\href }%
\providecommand \doibase [0]{http://dx.doi.org/}%
\providecommand \selectlanguage [0]{\@gobble}%
\providecommand \bibinfo  [0]{\@secondoftwo}%
\providecommand \bibfield  [0]{\@secondoftwo}%
\providecommand \translation [1]{[#1]}%
\providecommand \BibitemOpen [0]{}%
\providecommand \bibitemStop [0]{}%
\providecommand \bibitemNoStop [0]{.\EOS\space}%
\providecommand \EOS [0]{\spacefactor3000\relax}%
\providecommand \BibitemShut  [1]{\csname bibitem#1\endcsname}%
\let\auto@bib@innerbib\@empty
\bibitem [{\citenamefont {Islam}\ \emph {et~al.}(2022)\citenamefont {Islam},
  \citenamefont {Field}, \citenamefont {Hughes}, \citenamefont {Khanna},
  \citenamefont {Varma}, \citenamefont {Giesler}, \citenamefont {Scheel},
  \citenamefont {Kidder},\ and\ \citenamefont {Pfeiffer}}]{Islam:2022laz}%
  \BibitemOpen
  \bibfield  {author} {\bibinfo {author} {\bibfnamefont {Tousif}\ \bibnamefont
  {Islam}}, \bibinfo {author} {\bibfnamefont {Scott~E.}\ \bibnamefont {Field}},
  \bibinfo {author} {\bibfnamefont {Scott~A.}\ \bibnamefont {Hughes}}, \bibinfo
  {author} {\bibfnamefont {Gaurav}\ \bibnamefont {Khanna}}, \bibinfo {author}
  {\bibfnamefont {Vijay}\ \bibnamefont {Varma}}, \bibinfo {author}
  {\bibfnamefont {Matthew}\ \bibnamefont {Giesler}}, \bibinfo {author}
  {\bibfnamefont {Mark~A.}\ \bibnamefont {Scheel}}, \bibinfo {author}
  {\bibfnamefont {Lawrence~E.}\ \bibnamefont {Kidder}}, \ and\ \bibinfo
  {author} {\bibfnamefont {Harald~P.}\ \bibnamefont {Pfeiffer}},\ }\bibfield
  {title} {\enquote {\bibinfo {title} {{Surrogate model for gravitational wave
  signals from nonspinning, comparable-to large-mass-ratio black hole binaries
  built on black hole perturbation theory waveforms calibrated to numerical
  relativity}},}\ }\href {\doibase 10.1103/PhysRevD.106.104025} {\bibfield
  {journal} {\bibinfo  {journal} {Phys. Rev. D}\ }\textbf {\bibinfo {volume}
  {106}},\ \bibinfo {pages} {104025} (\bibinfo {year} {2022})},\ \Eprint
  {http://arxiv.org/abs/2204.01972} {arXiv:2204.01972 [gr-qc]} \BibitemShut
  {NoStop}%
\bibitem [{\citenamefont {Mroue}\ \emph {et~al.}(2013)\citenamefont {Mroue}
  \emph {et~al.}}]{Mroue:2013xna}%
  \BibitemOpen
  \bibfield  {author} {\bibinfo {author} {\bibfnamefont {Abdul~H.}\
  \bibnamefont {Mroue}} \emph {et~al.},\ }\bibfield  {title} {\enquote
  {\bibinfo {title} {{Catalog of 174 Binary Black Hole Simulations for
  Gravitational Wave Astronomy}},}\ }\href {\doibase
  10.1103/PhysRevLett.111.241104} {\bibfield  {journal} {\bibinfo  {journal}
  {Phys. Rev. Lett.}\ }\textbf {\bibinfo {volume} {111}},\ \bibinfo {pages}
  {241104} (\bibinfo {year} {2013})},\ \Eprint {http://arxiv.org/abs/1304.6077}
  {arXiv:1304.6077 [gr-qc]} \BibitemShut {NoStop}%
\bibitem [{\citenamefont {Boyle}\ \emph {et~al.}(2019)\citenamefont {Boyle}
  \emph {et~al.}}]{Boyle:2019kee}%
  \BibitemOpen
  \bibfield  {author} {\bibinfo {author} {\bibfnamefont {Michael}\ \bibnamefont
  {Boyle}} \emph {et~al.},\ }\bibfield  {title} {\enquote {\bibinfo {title}
  {{The SXS Collaboration catalog of binary black hole simulations}},}\ }\href
  {\doibase 10.1088/1361-6382/ab34e2} {\bibfield  {journal} {\bibinfo
  {journal} {Class. Quant. Grav.}\ }\textbf {\bibinfo {volume} {36}},\ \bibinfo
  {pages} {195006} (\bibinfo {year} {2019})},\ \Eprint
  {http://arxiv.org/abs/1904.04831} {arXiv:1904.04831 [gr-qc]} \BibitemShut
  {NoStop}%
\bibitem [{\citenamefont {Healy}\ \emph {et~al.}(2017)\citenamefont {Healy},
  \citenamefont {Lousto}, \citenamefont {Zlochower},\ and\ \citenamefont
  {Campanelli}}]{Healy:2017psd}%
  \BibitemOpen
  \bibfield  {author} {\bibinfo {author} {\bibfnamefont {James}\ \bibnamefont
  {Healy}}, \bibinfo {author} {\bibfnamefont {Carlos~O.}\ \bibnamefont
  {Lousto}}, \bibinfo {author} {\bibfnamefont {Yosef}\ \bibnamefont
  {Zlochower}}, \ and\ \bibinfo {author} {\bibfnamefont {Manuela}\ \bibnamefont
  {Campanelli}},\ }\bibfield  {title} {\enquote {\bibinfo {title} {{The RIT
  binary black hole simulations catalog}},}\ }\href {\doibase
  10.1088/1361-6382/aa91b1} {\bibfield  {journal} {\bibinfo  {journal} {Class.
  Quant. Grav.}\ }\textbf {\bibinfo {volume} {34}},\ \bibinfo {pages} {224001}
  (\bibinfo {year} {2017})},\ \Eprint {http://arxiv.org/abs/1703.03423}
  {arXiv:1703.03423 [gr-qc]} \BibitemShut {NoStop}%
\bibitem [{\citenamefont {Healy}\ \emph {et~al.}(2019)\citenamefont {Healy},
  \citenamefont {Lousto}, \citenamefont {Lange}, \citenamefont {O'Shaughnessy},
  \citenamefont {Zlochower},\ and\ \citenamefont {Campanelli}}]{Healy:2019jyf}%
  \BibitemOpen
  \bibfield  {author} {\bibinfo {author} {\bibfnamefont {James}\ \bibnamefont
  {Healy}}, \bibinfo {author} {\bibfnamefont {Carlos~O.}\ \bibnamefont
  {Lousto}}, \bibinfo {author} {\bibfnamefont {Jacob}\ \bibnamefont {Lange}},
  \bibinfo {author} {\bibfnamefont {Richard}\ \bibnamefont {O'Shaughnessy}},
  \bibinfo {author} {\bibfnamefont {Yosef}\ \bibnamefont {Zlochower}}, \ and\
  \bibinfo {author} {\bibfnamefont {Manuela}\ \bibnamefont {Campanelli}},\
  }\bibfield  {title} {\enquote {\bibinfo {title} {{Second RIT binary black
  hole simulations catalog and its application to gravitational waves parameter
  estimation}},}\ }\href {\doibase 10.1103/PhysRevD.100.024021} {\bibfield
  {journal} {\bibinfo  {journal} {Phys. Rev. D}\ }\textbf {\bibinfo {volume}
  {100}},\ \bibinfo {pages} {024021} (\bibinfo {year} {2019})},\ \Eprint
  {http://arxiv.org/abs/1901.02553} {arXiv:1901.02553 [gr-qc]} \BibitemShut
  {NoStop}%
\bibitem [{\citenamefont {Healy}\ and\ \citenamefont
  {Lousto}(2020)}]{Healy:2020vre}%
  \BibitemOpen
  \bibfield  {author} {\bibinfo {author} {\bibfnamefont {James}\ \bibnamefont
  {Healy}}\ and\ \bibinfo {author} {\bibfnamefont {Carlos~O.}\ \bibnamefont
  {Lousto}},\ }\bibfield  {title} {\enquote {\bibinfo {title} {{Third RIT
  binary black hole simulations catalog}},}\ }\href {\doibase
  10.1103/PhysRevD.102.104018} {\bibfield  {journal} {\bibinfo  {journal}
  {Phys. Rev. D}\ }\textbf {\bibinfo {volume} {102}},\ \bibinfo {pages}
  {104018} (\bibinfo {year} {2020})},\ \Eprint
  {http://arxiv.org/abs/2007.07910} {arXiv:2007.07910 [gr-qc]} \BibitemShut
  {NoStop}%
\bibitem [{\citenamefont {Healy}\ and\ \citenamefont
  {Lousto}(2022)}]{Healy:2022wdn}%
  \BibitemOpen
  \bibfield  {author} {\bibinfo {author} {\bibfnamefont {James}\ \bibnamefont
  {Healy}}\ and\ \bibinfo {author} {\bibfnamefont {Carlos~O.}\ \bibnamefont
  {Lousto}},\ }\bibfield  {title} {\enquote {\bibinfo {title} {{Fourth RIT
  binary black hole simulations catalog: Extension to eccentric orbits}},}\
  }\href {\doibase 10.1103/PhysRevD.105.124010} {\bibfield  {journal} {\bibinfo
   {journal} {Phys. Rev. D}\ }\textbf {\bibinfo {volume} {105}},\ \bibinfo
  {pages} {124010} (\bibinfo {year} {2022})},\ \Eprint
  {http://arxiv.org/abs/2202.00018} {arXiv:2202.00018 [gr-qc]} \BibitemShut
  {NoStop}%
\bibitem [{\citenamefont {Jani}\ \emph {et~al.}(2016)\citenamefont {Jani},
  \citenamefont {Healy}, \citenamefont {Clark}, \citenamefont {London},
  \citenamefont {Laguna},\ and\ \citenamefont {Shoemaker}}]{Jani:2016wkt}%
  \BibitemOpen
  \bibfield  {author} {\bibinfo {author} {\bibfnamefont {Karan}\ \bibnamefont
  {Jani}}, \bibinfo {author} {\bibfnamefont {James}\ \bibnamefont {Healy}},
  \bibinfo {author} {\bibfnamefont {James~A.}\ \bibnamefont {Clark}}, \bibinfo
  {author} {\bibfnamefont {Lionel}\ \bibnamefont {London}}, \bibinfo {author}
  {\bibfnamefont {Pablo}\ \bibnamefont {Laguna}}, \ and\ \bibinfo {author}
  {\bibfnamefont {Deirdre}\ \bibnamefont {Shoemaker}},\ }\bibfield  {title}
  {\enquote {\bibinfo {title} {{Georgia Tech Catalog of Gravitational
  Waveforms}},}\ }\href {\doibase 10.1088/0264-9381/33/20/204001} {\bibfield
  {journal} {\bibinfo  {journal} {Class. Quant. Grav.}\ }\textbf {\bibinfo
  {volume} {33}},\ \bibinfo {pages} {204001} (\bibinfo {year} {2016})},\
  \Eprint {http://arxiv.org/abs/1605.03204} {arXiv:1605.03204 [gr-qc]}
  \BibitemShut {NoStop}%
\bibitem [{\citenamefont {Hamilton}\ \emph {et~al.}(2023)\citenamefont
  {Hamilton} \emph {et~al.}}]{Hamilton:2023qkv}%
  \BibitemOpen
  \bibfield  {author} {\bibinfo {author} {\bibfnamefont {Eleanor}\ \bibnamefont
  {Hamilton}} \emph {et~al.},\ }\bibfield  {title} {\enquote {\bibinfo {title}
  {{A catalogue of precessing black-hole-binary numerical-relativity
  simulations}},}\ }\href@noop {} {\  (\bibinfo {year} {2023})},\ \Eprint
  {http://arxiv.org/abs/2303.05419} {arXiv:2303.05419 [gr-qc]} \BibitemShut
  {NoStop}%
\bibitem [{\citenamefont {Sundararajan}\ \emph {et~al.}(2007)\citenamefont
  {Sundararajan}, \citenamefont {Khanna},\ and\ \citenamefont
  {Hughes}}]{Sundararajan:2007jg}%
  \BibitemOpen
  \bibfield  {author} {\bibinfo {author} {\bibfnamefont {Pranesh~A.}\
  \bibnamefont {Sundararajan}}, \bibinfo {author} {\bibfnamefont {Gaurav}\
  \bibnamefont {Khanna}}, \ and\ \bibinfo {author} {\bibfnamefont {Scott~A.}\
  \bibnamefont {Hughes}},\ }\bibfield  {title} {\enquote {\bibinfo {title}
  {{Towards adiabatic waveforms for inspiral into Kerr black holes. I. A New
  model of the source for the time domain perturbation equation}},}\ }\href
  {\doibase 10.1103/PhysRevD.76.104005} {\bibfield  {journal} {\bibinfo
  {journal} {Phys. Rev. D}\ }\textbf {\bibinfo {volume} {76}},\ \bibinfo
  {pages} {104005} (\bibinfo {year} {2007})},\ \Eprint
  {http://arxiv.org/abs/gr-qc/0703028} {arXiv:gr-qc/0703028} \BibitemShut
  {NoStop}%
\bibitem [{\citenamefont {Sundararajan}\ \emph {et~al.}(2008)\citenamefont
  {Sundararajan}, \citenamefont {Khanna}, \citenamefont {Hughes},\ and\
  \citenamefont {Drasco}}]{Sundararajan:2008zm}%
  \BibitemOpen
  \bibfield  {author} {\bibinfo {author} {\bibfnamefont {Pranesh~A.}\
  \bibnamefont {Sundararajan}}, \bibinfo {author} {\bibfnamefont {Gaurav}\
  \bibnamefont {Khanna}}, \bibinfo {author} {\bibfnamefont {Scott~A.}\
  \bibnamefont {Hughes}}, \ and\ \bibinfo {author} {\bibfnamefont {Steve}\
  \bibnamefont {Drasco}},\ }\bibfield  {title} {\enquote {\bibinfo {title}
  {{Towards adiabatic waveforms for inspiral into Kerr black holes: II.
  Dynamical sources and generic orbits}},}\ }\href {\doibase
  10.1103/PhysRevD.78.024022} {\bibfield  {journal} {\bibinfo  {journal} {Phys.
  Rev. D}\ }\textbf {\bibinfo {volume} {78}},\ \bibinfo {pages} {024022}
  (\bibinfo {year} {2008})},\ \Eprint {http://arxiv.org/abs/0803.0317}
  {arXiv:0803.0317 [gr-qc]} \BibitemShut {NoStop}%
\bibitem [{\citenamefont {Sundararajan}\ \emph {et~al.}(2010)\citenamefont
  {Sundararajan}, \citenamefont {Khanna},\ and\ \citenamefont
  {Hughes}}]{Sundararajan:2010sr}%
  \BibitemOpen
  \bibfield  {author} {\bibinfo {author} {\bibfnamefont {Pranesh~A.}\
  \bibnamefont {Sundararajan}}, \bibinfo {author} {\bibfnamefont {Gaurav}\
  \bibnamefont {Khanna}}, \ and\ \bibinfo {author} {\bibfnamefont {Scott~A.}\
  \bibnamefont {Hughes}},\ }\bibfield  {title} {\enquote {\bibinfo {title}
  {{Binary black hole merger gravitational waves and recoil in the large mass
  ratio limit}},}\ }\href {\doibase 10.1103/PhysRevD.81.104009} {\bibfield
  {journal} {\bibinfo  {journal} {Phys. Rev. D}\ }\textbf {\bibinfo {volume}
  {81}},\ \bibinfo {pages} {104009} (\bibinfo {year} {2010})},\ \Eprint
  {http://arxiv.org/abs/1003.0485} {arXiv:1003.0485 [gr-qc]} \BibitemShut
  {NoStop}%
\bibitem [{\citenamefont {Zenginoglu}\ and\ \citenamefont
  {Khanna}(2011)}]{Zenginoglu:2011zz}%
  \BibitemOpen
  \bibfield  {author} {\bibinfo {author} {\bibfnamefont {Anil}\ \bibnamefont
  {Zenginoglu}}\ and\ \bibinfo {author} {\bibfnamefont {Gaurav}\ \bibnamefont
  {Khanna}},\ }\bibfield  {title} {\enquote {\bibinfo {title} {{Null infinity
  waveforms from extreme-mass-ratio inspirals in Kerr spacetime}},}\ }\href
  {\doibase 10.1103/PhysRevX.1.021017} {\bibfield  {journal} {\bibinfo
  {journal} {Phys. Rev. X}\ }\textbf {\bibinfo {volume} {1}},\ \bibinfo {pages}
  {021017} (\bibinfo {year} {2011})},\ \Eprint {http://arxiv.org/abs/1108.1816}
  {arXiv:1108.1816 [gr-qc]} \BibitemShut {NoStop}%
\bibitem [{\citenamefont {Fujita}\ and\ \citenamefont
  {Tagoshi}(2004)}]{Fujita:2004rb}%
  \BibitemOpen
  \bibfield  {author} {\bibinfo {author} {\bibfnamefont {Ryuichi}\ \bibnamefont
  {Fujita}}\ and\ \bibinfo {author} {\bibfnamefont {Hideyuki}\ \bibnamefont
  {Tagoshi}},\ }\bibfield  {title} {\enquote {\bibinfo {title} {{New numerical
  methods to evaluate homogeneous solutions of the Teukolsky equation}},}\
  }\href {\doibase 10.1143/PTP.112.415} {\bibfield  {journal} {\bibinfo
  {journal} {Prog. Theor. Phys.}\ }\textbf {\bibinfo {volume} {112}},\ \bibinfo
  {pages} {415--450} (\bibinfo {year} {2004})},\ \Eprint
  {http://arxiv.org/abs/gr-qc/0410018} {arXiv:gr-qc/0410018} \BibitemShut
  {NoStop}%
\bibitem [{\citenamefont {Fujita}\ and\ \citenamefont
  {Tagoshi}(2005)}]{Fujita:2005kng}%
  \BibitemOpen
  \bibfield  {author} {\bibinfo {author} {\bibfnamefont {Ryuichi}\ \bibnamefont
  {Fujita}}\ and\ \bibinfo {author} {\bibfnamefont {Hideyuki}\ \bibnamefont
  {Tagoshi}},\ }\bibfield  {title} {\enquote {\bibinfo {title} {{New Numerical
  Methods to Evaluate Homogeneous Solutions of the Teukolsky Equation II.
  Solutions of the Continued Fraction Equation}},}\ }\href {\doibase
  10.1143/PTP.113.1165} {\bibfield  {journal} {\bibinfo  {journal} {Prog.
  Theor. Phys.}\ }\textbf {\bibinfo {volume} {113}},\ \bibinfo {pages}
  {1165--1182} (\bibinfo {year} {2005})},\ \Eprint
  {http://arxiv.org/abs/0904.3818} {arXiv:0904.3818 [gr-qc]} \BibitemShut
  {NoStop}%
\bibitem [{\citenamefont {Mano}\ \emph {et~al.}(1996)\citenamefont {Mano},
  \citenamefont {Suzuki},\ and\ \citenamefont {Takasugi}}]{Mano:1996vt}%
  \BibitemOpen
  \bibfield  {author} {\bibinfo {author} {\bibfnamefont {Shuhei}\ \bibnamefont
  {Mano}}, \bibinfo {author} {\bibfnamefont {Hisao}\ \bibnamefont {Suzuki}}, \
  and\ \bibinfo {author} {\bibfnamefont {Eiichi}\ \bibnamefont {Takasugi}},\
  }\bibfield  {title} {\enquote {\bibinfo {title} {{Analytic solutions of the
  Teukolsky equation and their low frequency expansions}},}\ }\href {\doibase
  10.1143/PTP.95.1079} {\bibfield  {journal} {\bibinfo  {journal} {Prog. Theor.
  Phys.}\ }\textbf {\bibinfo {volume} {95}},\ \bibinfo {pages} {1079--1096}
  (\bibinfo {year} {1996})},\ \Eprint {http://arxiv.org/abs/gr-qc/9603020}
  {arXiv:gr-qc/9603020} \BibitemShut {NoStop}%
\bibitem [{\citenamefont {Throwe}(2010)}]{throwe2010high}%
  \BibitemOpen
  \bibfield  {author} {\bibinfo {author} {\bibfnamefont {William
  William~Thomas}\ \bibnamefont {Throwe}},\ }\emph {\bibinfo {title} {High
  precision calculation of generic extreme mass ratio inspirals}},\ \href@noop
  {} {Ph.D. thesis},\ \bibinfo  {school} {Massachusetts Institute of
  Technology} (\bibinfo {year} {2010})\BibitemShut {NoStop}%
\bibitem [{\citenamefont {O'Sullivan}\ and\ \citenamefont
  {Hughes}(2014)}]{OSullivan:2014ywd}%
  \BibitemOpen
  \bibfield  {author} {\bibinfo {author} {\bibfnamefont {Stephen}\ \bibnamefont
  {O'Sullivan}}\ and\ \bibinfo {author} {\bibfnamefont {Scott~A.}\ \bibnamefont
  {Hughes}},\ }\bibfield  {title} {\enquote {\bibinfo {title} {{Strong-field
  tidal distortions of rotating black holes: Formalism and results for
  circular, equatorial orbits}},}\ }\href {\doibase 10.1103/PhysRevD.91.109901}
  {\bibfield  {journal} {\bibinfo  {journal} {Phys. Rev. D}\ }\textbf {\bibinfo
  {volume} {90}},\ \bibinfo {pages} {124039} (\bibinfo {year} {2014})},\
  \bibinfo {note} {[Erratum: Phys.Rev.D 91, 109901 (2015)]},\ \Eprint
  {http://arxiv.org/abs/1407.6983} {arXiv:1407.6983 [gr-qc]} \BibitemShut
  {NoStop}%
\bibitem [{\citenamefont {Drasco}\ and\ \citenamefont
  {Hughes}(2006)}]{Drasco:2005kz}%
  \BibitemOpen
  \bibfield  {author} {\bibinfo {author} {\bibfnamefont {Steve}\ \bibnamefont
  {Drasco}}\ and\ \bibinfo {author} {\bibfnamefont {Scott~A.}\ \bibnamefont
  {Hughes}},\ }\bibfield  {title} {\enquote {\bibinfo {title} {{Gravitational
  wave snapshots of generic extreme mass ratio inspirals}},}\ }\href {\doibase
  10.1103/PhysRevD.73.024027} {\bibfield  {journal} {\bibinfo  {journal} {Phys.
  Rev. D}\ }\textbf {\bibinfo {volume} {73}},\ \bibinfo {pages} {024027}
  (\bibinfo {year} {2006})},\ \bibinfo {note} {[Erratum: Phys.Rev.D 88, 109905
  (2013), Erratum: Phys.Rev.D 90, 109905 (2014)]},\ \Eprint
  {http://arxiv.org/abs/gr-qc/0509101} {arXiv:gr-qc/0509101} \BibitemShut
  {NoStop}%
\bibitem [{\citenamefont {Lousto}\ \emph
  {et~al.}(2010{\natexlab{a}})\citenamefont {Lousto}, \citenamefont {Nakano},
  \citenamefont {Zlochower},\ and\ \citenamefont {Campanelli}}]{Lousto:2010tb}%
  \BibitemOpen
  \bibfield  {author} {\bibinfo {author} {\bibfnamefont {Carlos~O.}\
  \bibnamefont {Lousto}}, \bibinfo {author} {\bibfnamefont {Hiroyuki}\
  \bibnamefont {Nakano}}, \bibinfo {author} {\bibfnamefont {Yosef}\
  \bibnamefont {Zlochower}}, \ and\ \bibinfo {author} {\bibfnamefont {Manuela}\
  \bibnamefont {Campanelli}},\ }\bibfield  {title} {\enquote {\bibinfo {title}
  {{Intermediate Mass Ratio Black Hole Binaries: Numerical Relativity meets
  Perturbation Theory}},}\ }\href {\doibase 10.1103/PhysRevLett.104.211101}
  {\bibfield  {journal} {\bibinfo  {journal} {Phys. Rev. Lett.}\ }\textbf
  {\bibinfo {volume} {104}},\ \bibinfo {pages} {211101} (\bibinfo {year}
  {2010}{\natexlab{a}})},\ \Eprint {http://arxiv.org/abs/1001.2316}
  {arXiv:1001.2316 [gr-qc]} \BibitemShut {NoStop}%
\bibitem [{\citenamefont {Lousto}\ \emph
  {et~al.}(2010{\natexlab{b}})\citenamefont {Lousto}, \citenamefont {Nakano},
  \citenamefont {Zlochower},\ and\ \citenamefont {Campanelli}}]{Lousto:2010qx}%
  \BibitemOpen
  \bibfield  {author} {\bibinfo {author} {\bibfnamefont {Carlos~O.}\
  \bibnamefont {Lousto}}, \bibinfo {author} {\bibfnamefont {Hiroyuki}\
  \bibnamefont {Nakano}}, \bibinfo {author} {\bibfnamefont {Yosef}\
  \bibnamefont {Zlochower}}, \ and\ \bibinfo {author} {\bibfnamefont {Manuela}\
  \bibnamefont {Campanelli}},\ }\bibfield  {title} {\enquote {\bibinfo {title}
  {{Intermediate-mass-ratio black hole binaries: Intertwining numerical and
  perturbative techniques}},}\ }\href {\doibase 10.1103/PhysRevD.82.104057}
  {\bibfield  {journal} {\bibinfo  {journal} {Phys. Rev. D}\ }\textbf {\bibinfo
  {volume} {82}},\ \bibinfo {pages} {104057} (\bibinfo {year}
  {2010}{\natexlab{b}})},\ \Eprint {http://arxiv.org/abs/1008.4360}
  {arXiv:1008.4360 [gr-qc]} \BibitemShut {NoStop}%
\bibitem [{\citenamefont {Nakano}\ \emph {et~al.}(2011)\citenamefont {Nakano},
  \citenamefont {Zlochower}, \citenamefont {Lousto},\ and\ \citenamefont
  {Campanelli}}]{Nakano:2011pb}%
  \BibitemOpen
  \bibfield  {author} {\bibinfo {author} {\bibfnamefont {Hiroyuki}\
  \bibnamefont {Nakano}}, \bibinfo {author} {\bibfnamefont {Yosef}\
  \bibnamefont {Zlochower}}, \bibinfo {author} {\bibfnamefont {Carlos~O.}\
  \bibnamefont {Lousto}}, \ and\ \bibinfo {author} {\bibfnamefont {Manuela}\
  \bibnamefont {Campanelli}},\ }\bibfield  {title} {\enquote {\bibinfo {title}
  {{Intermediate-mass-ratio black hole binaries II: Modeling Trajectories and
  Gravitational Waveforms}},}\ }\href {\doibase 10.1103/PhysRevD.84.124006}
  {\bibfield  {journal} {\bibinfo  {journal} {Phys. Rev. D}\ }\textbf {\bibinfo
  {volume} {84}},\ \bibinfo {pages} {124006} (\bibinfo {year} {2011})},\
  \Eprint {http://arxiv.org/abs/1108.4421} {arXiv:1108.4421 [gr-qc]}
  \BibitemShut {NoStop}%
\bibitem [{\citenamefont {Navarro~Albalat}\ \emph {et~al.}(2023)\citenamefont
  {Navarro~Albalat}, \citenamefont {Zimmerman}, \citenamefont {Giesler},\ and\
  \citenamefont {Scheel}}]{NavarroAlbalat:2022tvh}%
  \BibitemOpen
  \bibfield  {author} {\bibinfo {author} {\bibfnamefont {Sergi}\ \bibnamefont
  {Navarro~Albalat}}, \bibinfo {author} {\bibfnamefont {Aaron}\ \bibnamefont
  {Zimmerman}}, \bibinfo {author} {\bibfnamefont {Matthew}\ \bibnamefont
  {Giesler}}, \ and\ \bibinfo {author} {\bibfnamefont {Mark~A.}\ \bibnamefont
  {Scheel}},\ }\bibfield  {title} {\enquote {\bibinfo {title} {{Success of the
  small mass-ratio approximation during the final orbits of binary black hole
  simulations}},}\ }\href {\doibase 10.1103/PhysRevD.107.084021} {\bibfield
  {journal} {\bibinfo  {journal} {Phys. Rev. D}\ }\textbf {\bibinfo {volume}
  {107}},\ \bibinfo {pages} {084021} (\bibinfo {year} {2023})},\ \Eprint
  {http://arxiv.org/abs/2207.04066} {arXiv:2207.04066 [gr-qc]} \BibitemShut
  {NoStop}%
\bibitem [{\citenamefont {Albalat}\ \emph {et~al.}(2022)\citenamefont
  {Albalat}, \citenamefont {Zimmerman}, \citenamefont {Giesler},\ and\
  \citenamefont {Scheel}}]{Albalat:2022lfz}%
  \BibitemOpen
  \bibfield  {author} {\bibinfo {author} {\bibfnamefont {Sergi~Navarro}\
  \bibnamefont {Albalat}}, \bibinfo {author} {\bibfnamefont {Aaron}\
  \bibnamefont {Zimmerman}}, \bibinfo {author} {\bibfnamefont {Matthew}\
  \bibnamefont {Giesler}}, \ and\ \bibinfo {author} {\bibfnamefont {Mark~A.}\
  \bibnamefont {Scheel}},\ }\bibfield  {title} {\enquote {\bibinfo {title}
  {{Redshift factor and the small mass-ratio limit in binary black hole
  simulations}},}\ }\href {\doibase 10.1103/PhysRevD.106.044006} {\bibfield
  {journal} {\bibinfo  {journal} {Phys. Rev. D}\ }\textbf {\bibinfo {volume}
  {106}},\ \bibinfo {pages} {044006} (\bibinfo {year} {2022})},\ \Eprint
  {http://arxiv.org/abs/2203.04893} {arXiv:2203.04893 [gr-qc]} \BibitemShut
  {NoStop}%
\bibitem [{\citenamefont {Ramos-Buades}\ \emph {et~al.}(2022)\citenamefont
  {Ramos-Buades}, \citenamefont {van~de Meent}, \citenamefont {Pfeiffer},
  \citenamefont {R\"uter}, \citenamefont {Scheel}, \citenamefont {Boyle},\ and\
  \citenamefont {Kidder}}]{Ramos-Buades:2022lgf}%
  \BibitemOpen
  \bibfield  {author} {\bibinfo {author} {\bibfnamefont {Antoni}\ \bibnamefont
  {Ramos-Buades}}, \bibinfo {author} {\bibfnamefont {Maarten}\ \bibnamefont
  {van~de Meent}}, \bibinfo {author} {\bibfnamefont {Harald~P.}\ \bibnamefont
  {Pfeiffer}}, \bibinfo {author} {\bibfnamefont {Hannes~R.}\ \bibnamefont
  {R\"uter}}, \bibinfo {author} {\bibfnamefont {Mark~A.}\ \bibnamefont
  {Scheel}}, \bibinfo {author} {\bibfnamefont {Michael}\ \bibnamefont {Boyle}},
  \ and\ \bibinfo {author} {\bibfnamefont {Lawrence~E.}\ \bibnamefont
  {Kidder}},\ }\bibfield  {title} {\enquote {\bibinfo {title} {{Eccentric
  binary black holes: Comparing numerical relativity and small mass-ratio
  perturbation theory}},}\ }\href {\doibase 10.1103/PhysRevD.106.124040}
  {\bibfield  {journal} {\bibinfo  {journal} {Phys. Rev. D}\ }\textbf {\bibinfo
  {volume} {106}},\ \bibinfo {pages} {124040} (\bibinfo {year} {2022})},\
  \Eprint {http://arxiv.org/abs/2209.03390} {arXiv:2209.03390 [gr-qc]}
  \BibitemShut {NoStop}%
\bibitem [{\citenamefont {van~de Meent}\ and\ \citenamefont
  {Pfeiffer}(2020)}]{vandeMeent:2020xgc}%
  \BibitemOpen
  \bibfield  {author} {\bibinfo {author} {\bibfnamefont {Maarten}\ \bibnamefont
  {van~de Meent}}\ and\ \bibinfo {author} {\bibfnamefont {Harald~P.}\
  \bibnamefont {Pfeiffer}},\ }\bibfield  {title} {\enquote {\bibinfo {title}
  {{Intermediate mass-ratio black hole binaries: Applicability of small
  mass-ratio perturbation theory}},}\ }\href {\doibase
  10.1103/PhysRevLett.125.181101} {\bibfield  {journal} {\bibinfo  {journal}
  {Phys. Rev. Lett.}\ }\textbf {\bibinfo {volume} {125}},\ \bibinfo {pages}
  {181101} (\bibinfo {year} {2020})},\ \Eprint
  {http://arxiv.org/abs/2006.12036} {arXiv:2006.12036 [gr-qc]} \BibitemShut
  {NoStop}%
\bibitem [{\citenamefont {Le~Tiec}(2014)}]{LeTiec:2014oez}%
  \BibitemOpen
  \bibfield  {author} {\bibinfo {author} {\bibfnamefont {Alexandre}\
  \bibnamefont {Le~Tiec}},\ }\bibfield  {title} {\enquote {\bibinfo {title}
  {{The Overlap of Numerical Relativity, Perturbation Theory and Post-Newtonian
  Theory in the Binary Black Hole Problem}},}\ }\href {\doibase
  10.1142/S0218271814300225} {\bibfield  {journal} {\bibinfo  {journal} {Int.
  J. Mod. Phys. D}\ }\textbf {\bibinfo {volume} {23}},\ \bibinfo {pages}
  {1430022} (\bibinfo {year} {2014})},\ \Eprint
  {http://arxiv.org/abs/1408.5505} {arXiv:1408.5505 [gr-qc]} \BibitemShut
  {NoStop}%
\bibitem [{\citenamefont {Le~Tiec}\ \emph {et~al.}(2013)\citenamefont {Le~Tiec}
  \emph {et~al.}}]{LeTiec:2013uey}%
  \BibitemOpen
  \bibfield  {author} {\bibinfo {author} {\bibfnamefont {Alexandre}\
  \bibnamefont {Le~Tiec}} \emph {et~al.},\ }\bibfield  {title} {\enquote
  {\bibinfo {title} {{Periastron Advance in Spinning Black Hole Binaries:
  Gravitational Self-Force from Numerical Relativity}},}\ }\href {\doibase
  10.1103/PhysRevD.88.124027} {\bibfield  {journal} {\bibinfo  {journal} {Phys.
  Rev. D}\ }\textbf {\bibinfo {volume} {88}},\ \bibinfo {pages} {124027}
  (\bibinfo {year} {2013})},\ \Eprint {http://arxiv.org/abs/1309.0541}
  {arXiv:1309.0541 [gr-qc]} \BibitemShut {NoStop}%
\bibitem [{\citenamefont {Le~Tiec}\ \emph {et~al.}(2012)\citenamefont
  {Le~Tiec}, \citenamefont {Barausse},\ and\ \citenamefont
  {Buonanno}}]{LeTiec:2011dp}%
  \BibitemOpen
  \bibfield  {author} {\bibinfo {author} {\bibfnamefont {Alexandre}\
  \bibnamefont {Le~Tiec}}, \bibinfo {author} {\bibfnamefont {Enrico}\
  \bibnamefont {Barausse}}, \ and\ \bibinfo {author} {\bibfnamefont
  {Alessandra}\ \bibnamefont {Buonanno}},\ }\bibfield  {title} {\enquote
  {\bibinfo {title} {{Gravitational Self-Force Correction to the Binding Energy
  of Compact Binary Systems}},}\ }\href {\doibase
  10.1103/PhysRevLett.108.131103} {\bibfield  {journal} {\bibinfo  {journal}
  {Phys. Rev. Lett.}\ }\textbf {\bibinfo {volume} {108}},\ \bibinfo {pages}
  {131103} (\bibinfo {year} {2012})},\ \Eprint {http://arxiv.org/abs/1111.5609}
  {arXiv:1111.5609 [gr-qc]} \BibitemShut {NoStop}%
\bibitem [{\citenamefont {Le~Tiec}(2011)}]{LeTiec:2011ru}%
  \BibitemOpen
  \bibfield  {author} {\bibinfo {author} {\bibfnamefont {Alexandre}\
  \bibnamefont {Le~Tiec}},\ }\bibfield  {title} {\enquote {\bibinfo {title}
  {{Perturbative, Post-Newtonian, and General Relativistic Dynamics of Black
  Hole Binaries}},}\ }in\ \href@noop {} {\emph {\bibinfo {booktitle} {{46th
  Rencontres de Moriond on Gravitational Waves and Experimental Gravity}}}}\
  (\bibinfo {year} {2011})\ pp.\ \bibinfo {pages} {81--84},\ \Eprint
  {http://arxiv.org/abs/1109.6848} {arXiv:1109.6848 [gr-qc]} \BibitemShut
  {NoStop}%
\bibitem [{\citenamefont {Pound}\ and\ \citenamefont
  {Wardell}(2021)}]{Pound:2021qin}%
  \BibitemOpen
  \bibfield  {author} {\bibinfo {author} {\bibfnamefont {Adam}\ \bibnamefont
  {Pound}}\ and\ \bibinfo {author} {\bibfnamefont {Barry}\ \bibnamefont
  {Wardell}},\ }\bibfield  {title} {\enquote {\bibinfo {title} {{Black hole
  perturbation theory and gravitational self-force}},}\ }\href@noop {} {\
  (\bibinfo {year} {2021})},\ \Eprint {http://arxiv.org/abs/2101.04592}
  {arXiv:2101.04592 [gr-qc]} \BibitemShut {NoStop}%
\bibitem [{\citenamefont {Miller}\ and\ \citenamefont
  {Pound}(2021)}]{Miller:2020bft}%
  \BibitemOpen
  \bibfield  {author} {\bibinfo {author} {\bibfnamefont {Jeremy}\ \bibnamefont
  {Miller}}\ and\ \bibinfo {author} {\bibfnamefont {Adam}\ \bibnamefont
  {Pound}},\ }\bibfield  {title} {\enquote {\bibinfo {title} {{Two-timescale
  evolution of extreme-mass-ratio inspirals: waveform generation scheme for
  quasicircular orbits in Schwarzschild spacetime}},}\ }\href {\doibase
  10.1103/PhysRevD.103.064048} {\bibfield  {journal} {\bibinfo  {journal}
  {Phys. Rev. D}\ }\textbf {\bibinfo {volume} {103}},\ \bibinfo {pages}
  {064048} (\bibinfo {year} {2021})},\ \Eprint
  {http://arxiv.org/abs/2006.11263} {arXiv:2006.11263 [gr-qc]} \BibitemShut
  {NoStop}%
\bibitem [{\citenamefont {Wardell}\ \emph {et~al.}(2021)\citenamefont
  {Wardell}, \citenamefont {Pound}, \citenamefont {Warburton}, \citenamefont
  {Miller}, \citenamefont {Durkan},\ and\ \citenamefont
  {Le~Tiec}}]{Wardell:2021fyy}%
  \BibitemOpen
  \bibfield  {author} {\bibinfo {author} {\bibfnamefont {Barry}\ \bibnamefont
  {Wardell}}, \bibinfo {author} {\bibfnamefont {Adam}\ \bibnamefont {Pound}},
  \bibinfo {author} {\bibfnamefont {Niels}\ \bibnamefont {Warburton}}, \bibinfo
  {author} {\bibfnamefont {Jeremy}\ \bibnamefont {Miller}}, \bibinfo {author}
  {\bibfnamefont {Leanne}\ \bibnamefont {Durkan}}, \ and\ \bibinfo {author}
  {\bibfnamefont {Alexandre}\ \bibnamefont {Le~Tiec}},\ }\bibfield  {title}
  {\enquote {\bibinfo {title} {{Gravitational waveforms for compact binaries
  from second-order self-force theory}},}\ }\href@noop {} {\  (\bibinfo {year}
  {2021})},\ \Eprint {http://arxiv.org/abs/2112.12265} {arXiv:2112.12265
  [gr-qc]} \BibitemShut {NoStop}%
\bibitem [{\citenamefont {Rifat}\ \emph {et~al.}(2020)\citenamefont {Rifat},
  \citenamefont {Field}, \citenamefont {Khanna},\ and\ \citenamefont
  {Varma}}]{Rifat:2019ltp}%
  \BibitemOpen
  \bibfield  {author} {\bibinfo {author} {\bibfnamefont {Nur E.~M.}\
  \bibnamefont {Rifat}}, \bibinfo {author} {\bibfnamefont {Scott~E.}\
  \bibnamefont {Field}}, \bibinfo {author} {\bibfnamefont {Gaurav}\
  \bibnamefont {Khanna}}, \ and\ \bibinfo {author} {\bibfnamefont {Vijay}\
  \bibnamefont {Varma}},\ }\bibfield  {title} {\enquote {\bibinfo {title}
  {{Surrogate model for gravitational wave signals from comparable and
  large-mass-ratio black hole binaries}},}\ }\href {\doibase
  10.1103/PhysRevD.101.081502} {\bibfield  {journal} {\bibinfo  {journal}
  {Phys. Rev. D}\ }\textbf {\bibinfo {volume} {101}},\ \bibinfo {pages}
  {081502} (\bibinfo {year} {2020})},\ \Eprint
  {http://arxiv.org/abs/1910.10473} {arXiv:1910.10473 [gr-qc]} \BibitemShut
  {NoStop}%
\bibitem [{\citenamefont {Islam}\ \emph {et~al.}(2023)\citenamefont {Islam},
  \citenamefont {Field},\ and\ \citenamefont {Khanna}}]{Islam:2023mob}%
  \BibitemOpen
  \bibfield  {author} {\bibinfo {author} {\bibfnamefont {Tousif}\ \bibnamefont
  {Islam}}, \bibinfo {author} {\bibfnamefont {Scott~E.}\ \bibnamefont {Field}},
  \ and\ \bibinfo {author} {\bibfnamefont {Gaurav}\ \bibnamefont {Khanna}},\
  }\bibfield  {title} {\enquote {\bibinfo {title} {{Remnant black hole
  properties from numerical-relativity-informed perturbation theory and
  implications for waveform modelling}},}\ }\href@noop {} {\  (\bibinfo {year}
  {2023})},\ \Eprint {http://arxiv.org/abs/2301.07215} {arXiv:2301.07215
  [gr-qc]} \BibitemShut {NoStop}%
\bibitem [{\citenamefont {Islam}(2023)}]{Islam:2023qyt}%
  \BibitemOpen
  \bibfield  {author} {\bibinfo {author} {\bibfnamefont {Tousif}\ \bibnamefont
  {Islam}},\ }\bibfield  {title} {\enquote {\bibinfo {title} {{Interplay
  between numerical-relativity and black hole perturbation theory in the
  intermediate-mass-ratio regime}},}\ }\href@noop {} {\  (\bibinfo {year}
  {2023})},\ \Eprint {http://arxiv.org/abs/2306.08771} {arXiv:2306.08771
  [gr-qc]} \BibitemShut {NoStop}%
\bibitem [{\citenamefont {Islam}\ and\ \citenamefont
  {Khanna}(2023{\natexlab{a}})}]{Islam:2023aec}%
  \BibitemOpen
  \bibfield  {author} {\bibinfo {author} {\bibfnamefont {Tousif}\ \bibnamefont
  {Islam}}\ and\ \bibinfo {author} {\bibfnamefont {Gaurav}\ \bibnamefont
  {Khanna}},\ }\bibfield  {title} {\enquote {\bibinfo {title} {{Interplay
  between numerical relativity and perturbation theory : finite size
  effects}},}\ }\href@noop {} {\  (\bibinfo {year} {2023}{\natexlab{a}})},\
  \Eprint {http://arxiv.org/abs/2306.08767} {arXiv:2306.08767 [gr-qc]}
  \BibitemShut {NoStop}%
\bibitem [{\citenamefont {Islam}\ and\ \citenamefont
  {Khanna}(2023{\natexlab{b}})}]{Islam:2023jak}%
  \BibitemOpen
  \bibfield  {author} {\bibinfo {author} {\bibfnamefont {Tousif}\ \bibnamefont
  {Islam}}\ and\ \bibinfo {author} {\bibfnamefont {Gaurav}\ \bibnamefont
  {Khanna}},\ }\bibfield  {title} {\enquote {\bibinfo {title} {{On the
  approximate relation between black-hole perturbation theory and numerical
  relativity}},}\ }\href@noop {} {\  (\bibinfo {year} {2023}{\natexlab{b}})},\
  \Eprint {http://arxiv.org/abs/2307.03155} {arXiv:2307.03155 [gr-qc]}
  \BibitemShut {NoStop}%
\bibitem [{\citenamefont {Lousto}\ and\ \citenamefont
  {Healy}(2020)}]{Lousto:2020tnb}%
  \BibitemOpen
  \bibfield  {author} {\bibinfo {author} {\bibfnamefont {Carlos~O.}\
  \bibnamefont {Lousto}}\ and\ \bibinfo {author} {\bibfnamefont {James}\
  \bibnamefont {Healy}},\ }\bibfield  {title} {\enquote {\bibinfo {title}
  {{Exploring the Small Mass Ratio Binary Black Hole Merger via
  Zeno\textquoteright{}s Dichotomy Approach}},}\ }\href {\doibase
  10.1103/PhysRevLett.125.191102} {\bibfield  {journal} {\bibinfo  {journal}
  {Phys. Rev. Lett.}\ }\textbf {\bibinfo {volume} {125}},\ \bibinfo {pages}
  {191102} (\bibinfo {year} {2020})},\ \Eprint
  {http://arxiv.org/abs/2006.04818} {arXiv:2006.04818 [gr-qc]} \BibitemShut
  {NoStop}%
\bibitem [{\citenamefont {Lousto}\ and\ \citenamefont
  {Healy}(2022)}]{Lousto:2022hoq}%
  \BibitemOpen
  \bibfield  {author} {\bibinfo {author} {\bibfnamefont {Carlos~O.}\
  \bibnamefont {Lousto}}\ and\ \bibinfo {author} {\bibfnamefont {James}\
  \bibnamefont {Healy}},\ }\bibfield  {title} {\enquote {\bibinfo {title}
  {{Study of the Intermediate Mass Ratio Black Hole Binary Merger up to 1000:1
  with Numerical Relativity}},}\ }\href@noop {} {\  (\bibinfo {year} {2022})},\
  \Eprint {http://arxiv.org/abs/2203.08831} {arXiv:2203.08831 [gr-qc]}
  \BibitemShut {NoStop}%
\bibitem [{\citenamefont {Field}\ \emph {et~al.}()\citenamefont {Field},
  \citenamefont {Islam}, \citenamefont {Khanna}, \citenamefont {Rifat},\ and\
  \citenamefont {Varma}}]{BHPTSurrogate}%
  \BibitemOpen
  \bibfield  {author} {\bibinfo {author} {\bibfnamefont {Scott}\ \bibnamefont
  {Field}}, \bibinfo {author} {\bibfnamefont {Tousif}\ \bibnamefont {Islam}},
  \bibinfo {author} {\bibfnamefont {Gaurav}\ \bibnamefont {Khanna}}, \bibinfo
  {author} {\bibfnamefont {Nur}\ \bibnamefont {Rifat}}, \ and\ \bibinfo
  {author} {\bibfnamefont {Vijay}\ \bibnamefont {Varma}},\ }\href@noop {}
  {\enquote {\bibinfo {title} {{BHPTNRSurrogate}},}\ }\bibinfo {note}
  {\url{http://bhptoolkit.org/BHPTNRSurrogate/}}\BibitemShut {NoStop}%
\bibitem [{BHP()}]{BHPToolkit}%
  \BibitemOpen
  \href@noop {} {\enquote {\bibinfo {title} {{Black Hole Perturbation
  Toolkit}},}\ }\bibinfo {howpublished}
  {(\href{http://bhptoolkit.org/}{bhptoolkit.org})}\BibitemShut {NoStop}%
\bibitem [{\citenamefont {Islam}\ \emph {et~al.}()\citenamefont {Islam},
  \citenamefont {Field},\ and\ \citenamefont {Khanna}}]{gwremnant}%
  \BibitemOpen
  \bibfield  {author} {\bibinfo {author} {\bibfnamefont {Tousif}\ \bibnamefont
  {Islam}}, \bibinfo {author} {\bibfnamefont {Scott}\ \bibnamefont {Field}}, \
  and\ \bibinfo {author} {\bibfnamefont {Gaurav}\ \bibnamefont {Khanna}},\
  }\href@noop {} {\enquote {\bibinfo {title} {{BHPTNRSurrogate}},}\ }\bibinfo
  {note} {\url{https://pypi.org/project/gw-remnant/}}\BibitemShut {NoStop}%
\bibitem [{sci()}]{scipyfit}%
  \BibitemOpen
  \href {\doibase
  https://docs.scipy.org/doc/scipy/reference/generated/scipy.optimize.curve\_fit.html}
  {\enquote {\bibinfo {title} {{scipy.optimize.cruve\_fit}},}\ }\BibitemShut
  {NoStop}%
\bibitem [{\citenamefont {Barausse}\ \emph {et~al.}(2021)\citenamefont
  {Barausse}, \citenamefont {Berti}, \citenamefont {Cardoso}, \citenamefont
  {Hughes},\ and\ \citenamefont {Khanna}}]{Barausse:2021}%
  \BibitemOpen
  \bibfield  {author} {\bibinfo {author} {\bibfnamefont {Enrico}\ \bibnamefont
  {Barausse}}, \bibinfo {author} {\bibfnamefont {Emanuele}\ \bibnamefont
  {Berti}}, \bibinfo {author} {\bibfnamefont {Vitor}\ \bibnamefont {Cardoso}},
  \bibinfo {author} {\bibfnamefont {Scott~A.}\ \bibnamefont {Hughes}}, \ and\
  \bibinfo {author} {\bibfnamefont {Gaurav}\ \bibnamefont {Khanna}},\
  }\bibfield  {title} {\enquote {\bibinfo {title} {Divergences in
  gravitational-wave emission and absorption from extreme mass ratio
  binaries},}\ }\href {\doibase 10.1103/PhysRevD.104.064031} {\bibfield
  {journal} {\bibinfo  {journal} {Phys. Rev. D}\ }\textbf {\bibinfo {volume}
  {104}},\ \bibinfo {pages} {064031} (\bibinfo {year} {2021})}\BibitemShut
  {NoStop}%
\end{thebibliography}%

\end{document}